\documentclass[useAMS,usenatbib]{mn2e}

\usepackage{epsfig}
\usepackage{amsmath}
\usepackage{amssymb}
\usepackage{mathrsfs}
\usepackage{graphicx}
\usepackage{relsize}
\usepackage{mathtools}
\usepackage{multirow}

\newcommand{\apj}{ApJ} 
\newcommand{\apjl}{ApJ} 
\newcommand{\apjs}{ApJ} 
\newcommand{\aap}{A\&A} 
\newcommand{\araa}{ARAA} 
\newcommand{\mnras}{MNRAS} 
\newcommand\aapr{A\&A~Rev.} 
          
\newcommand{\mach}{\mathcal{M}}
\newcommand{\machv}{\mathcal{M}_{V}}
\newcommand{\machm}{\mathcal{M}_{M}}

\newcommand{\sigs}{\sigma_s}
\newcommand{\sigsv}{\sigma_{s,V}}
\newcommand{\sigsm}{\sigma_{s,M}}
\newcommand{\kinj}{k_\mathrm{inj}}

\newcommand{\deriv}{\mathrm{d}}

\newcommand{\sfr}{\mathrm{SFR}}

\newcommand{\alphavir}{\alpha_\mathrm{vir}}

\newcommand{\tff}{t_\mathrm{ff}}
\newcommand{\scrit}{s_\mathrm{crit}}
\newcommand{\rhocrit}{\rho_\mathrm{crit}}

\newcommand{\vect}[1]{{\mathbf{#1}}}
\newcommand{\cs}{c_\mathrm{s}}
\newcommand{\csv}{c_{\mathrm{s},V}}
\newcommand{\csm}{c_{\mathrm{s},M}}
\newcommand{\cszero}{c_\mathrm{s,0}}
\newcommand{\csone}{c_\mathrm{s,1}}
\newcommand{\cstwo}{c_\mathrm{s,2}}
\newcommand{\va}{v_\mathrm{A}}
\newcommand{\cm}{\mbox{cm}}
\newcommand{\pc}{\mbox{pc}}

\title[The density PDF and SFR of polytropic turbulence]{The density structure and star formation rate of non-isothermal polytropic turbulence}

\author[Federrath \& Banerjee]{
Christoph~Federrath$^{1}$\thanks{E-mail: christoph.federrath@anu.edu.au} \&
Supratik~Banerjee$^{2}$\thanks{E-mail: supratik.banerjee@uni-koeln.de}\\
$^{1}$Research School of Astronomy and Astrophysics, The Australian National University, Canberra, ACT~2611, Australia\\
$^{2}$Institut f\"ur Geophysik und Meteorologie, Pohligstrasse~3, D-50969 Cologne, Germany
}

\begin{document}

\maketitle

\begin{abstract}
The interstellar medium of galaxies is governed by supersonic turbulence, which likely controls the star formation rate (SFR) and the initial mass function (IMF). Interstellar turbulence is non-universal, with a wide range of Mach numbers, magnetic fields strengths, and driving mechanisms. Although some of these parameters were explored, most previous works assumed that the gas is isothermal. However, we know that cold molecular clouds form out of the warm atomic medium, with the gas passing through chemical and thermodynamic phases that are not isothermal. Here we determine the role of temperature variations by modelling non-isothermal turbulence with a polytropic equation of state (EOS), where pressure and temperature are functions of gas density, $P\sim\rho^\Gamma$, $T\sim\rho^{\Gamma-1}$. We use grid resolutions of $2048^3$ cells and compare polytropic exponents $\Gamma=0.7$ (soft EOS), $\Gamma=1$ (isothermal EOS), and $\Gamma=5/3$ (stiff EOS). We find a complex network of non-isothermal filaments with more small-scale fragmentation occurring for $\Gamma<1$, while $\Gamma>1$ smoothes out density contrasts. The density probability distribution function (PDF) is significantly affected by temperature variations, with a power-law tail developing at low densities for $\Gamma>1$. In contrast, the PDF becomes closer to a lognormal distribution for $\Gamma\lesssim1$. We derive and test a new density variance -- Mach number relation that takes $\Gamma$ into account. This new relation is relevant for theoretical models of the SFR and IMF, because it determines the dense gas mass fraction of a cloud, from which stars form. We derive the SFR as a function of $\Gamma$ and find that it decreases by a factor of $\sim5$ from $\Gamma=0.7$ to $\Gamma=5/3$.
\end{abstract}


\section{Introduction}

Interstellar turbulence is a key for star formation \citep{MacLowKlessen2004,ElmegreenScalo2004,McKeeOstriker2007,PadoanEtAl2014}. Yet our observational and theoretical understanding of interstellar turbulence is limited. This is primarily because turbulence is an intrinsically complex, three-dimensional (3D) phenomenon occurring only at very high Reynolds numbers \citep{Krumholz2014}, which are difficult to achieve in terrestrial experiments. What we do know is that turbulence in the interstellar medium is highly compressible and supersonic \citep{Larson1981,HeyerBrunt2004,RomanDuvalEtAl2011,HennebelleFalgarone2012}, significantly exceeding the complexity of incompressible turbulence \citep{Kolmogorov1941c,Frisch1995}. Supersonic, compressible turbulence is difficult to study analytically, but some important steps have been taken \citep{LazarianPogosyan2000,BoldyrevNordlundPadoan2002,LazarianEsquivel2003,SchmidtFederrathKlessen2008,GaltierBanerjee2011,Aluie2011,Aluie2013,BanerjeeGaltier2013,BanerjeeGaltier2014}. In order to unravel the statistics and properties of turbulence in detail, however, one must ultimately resort to full 3D computer simulations.

Attempts to model supersonic turbulence in a computer reach back to the early studies by \citet{PorterPouquetWoodward1992} and \citet{PorterWoodward1994}. However, it is only within the last few years with the advent of supercomputers that we can now measure the scaling of the turbulent density and velocity with high precision \citep{ChoLazarian2003,KritsukEtAl2007,KowalLazarian2007,LemasterStone2009,SchmidtEtAl2009,FederrathDuvalKlessenSchmidtMacLow2010,KonstandinEtAl2012,Federrath2013}. But an important limitation of these studies is that they all rely on the assumption of isothermal gas. The real interstellar medium, however, consists of several density and temperature phases \citep{HollenbachWernerSalpeter1971,McKee1989,WolfireEtAl1995,Ferriere2001}. Even molecular clouds do exhibit potentially important temperature variations that can be approximated with a polytropic EOS, relating pressure $P$, temperature $T$ and density $n$,
\begin{equation} \label{eq:eosbasic}
P \sim n^\Gamma,\;T\sim n^{\Gamma-1}.
\end{equation}
The polytropic exponent $\Gamma$ is close to unity, $\Gamma=1$ (isothermal gas), over a wide range of densities, from hydrogen number densities of \mbox{$n\sim1$--$10^{10}\,\cm^{-3}$}, with the temperature varying in the range $3\,\mathrm{K} < T < 10\,\mathrm{K}$ for solar metallicity gas \citep{OmukaiEtAl2005}. Radiation-hydrodynamical calculations including chemical evolution and cooling by \citet{MasunagaInutsuka2000} also show that $\Gamma\sim1$ for $n\lesssim10^9\,\cm^{-3}$. More recent 3D calculations including a detailed chemical network find that \mbox{$\Gamma\sim0.5$--$0.9$} in the range $10\,\cm^{-3}\lesssim n \lesssim 10^4\,\cm^{-3}$ \citep{GloverMacLow2007a,GloverMacLow2007b}, followed by $\Gamma\sim1$ at higher densities \citep{GloverFederrathMacLowKlessen2010}. As the gas becomes optically thick, $\Gamma$ rises to $\Gamma\sim1.1$ for $10^9\lesssim n/\cm^{-3}\lesssim10^{11}$, $\Gamma\sim1.4$ for $10^{11}\lesssim n/\cm^{-3}\lesssim10^{16}$, followed by a phase where $\Gamma\sim1.1$ in which molecular hydrogen is dissociated ($10^{16}\lesssim n/\cm^{-3}\lesssim10^{21}$) \citep{MasunagaInutsuka2000}. Finally, the gas becomes almost completely optically thick ($\Gamma=5/3$), when a new star is born ($n\gtrsim10^{21}\,\cm^{-3}$). It must be emphasised that all the phases with $n\gtrsim10^{10}\,\cm^{-3}$ occur inside dense, collapsing cores with sizes $<0.1\,\pc$, while the phases with \mbox{$\Gamma\sim0.5$--$1.1$} are relevant for molecular cloud scales, \mbox{$L\sim0.1$--$100\,\pc$}. Turbulent gas in the early Universe likely had a somewhat higher effective polytropic exponent, because of slightly less efficient cooling \citep{AbelBryanNorman2002,GreifEtAl2008,WiseTurkAbel2008,SchleicherEtAl2010,RomeoBurkertAgertz2010,ClarkEtAl2011,HoffmannRomeo2012,SchoberEtAl2012,SafranekShraderEtAl2012,LatifEtAl2013}.

The aim of this study is to determine the role of temperature variations on the filamentary structure and the density PDF of molecular clouds in the interstellar medium. Here we measure the density PDF in non-isothermal gas governed by polytropic turbulence and derive the density variance -- Mach number relation as a function of the polytropic exponent $\Gamma$. Finally, we use the new density PDF to determine the SFR in polytropic clouds, given the virial parameter, turbulent driving, Mach number and polytropic $\Gamma$. We note that the influence of temperature variations has been explored in previous complementary studies \citep{Vazquez1994,PassotVazquez1998,WadaNorman2001,KritsukNorman2002,LiKlessenMacLow2003,JappsenEtAl2005,AuditHennebelle2005,AuditHennebelle2010,HennebelleAudit2007,KissmannEtAl2008,SeifriedEtAl2011,KimKimOstriker2011,PetersEtAl2012,GazolKim2013,TociGalli2015}, and we extend these here to much higher resolution and focus on the implications for star formation.

The paper is organised as follows. Section~\ref{sec:methods} summarises the hydrodynamical simulation methods. Sections~\ref{sec:structure} and~\ref{sec:tevol} present the filamentary structure and time evolution of polytropic turbulence. In Section~\ref{sec:pdfs} we determine the density PDF of polytropic clouds and we derive a new density variance -- Mach number relation for polytropic gases in Section~\ref{sec:densityvariancemachrelation}. We then show in Section~\ref{sec:sfr} that this new PDF leads to SFRs varying by a factor of $\sim5$ in the polytropic regime with $0.7 \leq \Gamma \leq 5/3$, occurring in real molecular clouds. Our conclusions are listed in Section~\ref{sec:conclusions}.

\section{Numerical simulations} \label{sec:methods}

\begin{table*}
\caption{Simulation parameters and statistical measures}
\label{tab:sims}
\def\arraystretch{0.2}
\setlength{\tabcolsep}{7.0pt}
\begin{tabular}{lrrrrrrrrr}
\hline
Simulation & $N_\mathrm{res}^3$ & $\Gamma$ & $\machv$ & $\machm$ & Sim $\sigsv$ & Sim $\sigsm$ & PDF $\sigsv$ & PDF $\sigsm$ & PDF $\theta$ \\
(1) & (2) & (3) & (4) & (5) & (6) & (7) & (8) & (9) & (10) \\
\hline
PT2048G0.7  &  $2048^3$  &  $0.7$  &  $8.4\pm0.4$  &  $12.4\pm0.5$  &  $1.83\pm0.08$  &  $1.74\pm0.07$  &  $1.73\pm0.03$  &  $1.56\pm0.05$  &  $0.07\pm0.02$ \\
PT1024G0.7  &  $1024^3$  &  $0.7$  &  $8.6\pm0.3$  &  $12.5\pm0.6$  &  $1.84\pm0.11$  &  $1.67\pm0.04$  &  $1.78\pm0.03$  &  $1.60\pm0.05$  &  $0.07\pm0.02$ \\
PT512G0.7    &  $512^3$    &  $0.7$  &  $8.5\pm0.4$  &  $12.6\pm0.6$  &  $1.84\pm0.10$  &  $1.67\pm0.05$  &  $1.69\pm0.09$  &  $1.53\pm0.16$  &  $0.07\pm0.03$ \\
PT256G0.7    &  $256^3$    &  $0.7$  &  $8.5\pm0.4$  &  $12.6\pm0.7$  &  $1.92\pm0.08$  &  $1.60\pm0.05$  &  $1.78\pm0.06$  &  $1.60\pm0.11$  &  $0.07\pm0.04$ \\
\hline
PT1024G1     &  $1024^3$  &  $1$  &  $11.6\pm0.6$  &  $10.7\pm0.6$  &  $1.94\pm0.11$  &  $1.57\pm0.07$  &  $1.82\pm0.08$  &  $1.59\pm0.11$  &  $0.10\pm0.04$ \\
PT512G1       &  $512^3$    &  $1$  &  $11.8\pm0.6$  &  $10.7\pm0.6$  &  $1.98\pm0.11$  &  $1.57\pm0.06$  &  $1.84\pm0.07$  &  $1.59\pm0.11$  &  $0.10\pm0.04$ \\
PT256G1       &  $256^3$    &  $1$  &  $11.7\pm0.6$  &  $10.7\pm0.6$  &  $2.05\pm0.13$  &  $1.51\pm0.05$  &  $2.01\pm0.12$  &  $1.52\pm0.17$  &  $0.20\pm0.07$ \\
\hline
PT2048G5/3  &  $2048^3$  &  $5/3$  &  $13.3\pm0.5$  &  $6.3\pm0.4$  &  $2.17\pm0.11$  &  $1.27\pm0.03$  &  $2.12\pm0.15$  &  $1.26\pm0.17$  &  $0.41\pm0.08$ \\
PT1024G5/3  &  $1024^3$  &  $5/3$  &  $13.2\pm0.5$  &  $6.3\pm0.3$  &  $2.17\pm0.15$  &  $1.27\pm0.04$  &  $2.10\pm0.16$  &  $1.26\pm0.19$  &  $0.40\pm0.09$ \\
PT512G5/3    &  $512^3$    &  $5/3$  &  $13.2\pm0.5$  &  $6.3\pm0.3$  &  $2.12\pm0.15$  &  $1.28\pm0.04$  &  $2.04\pm0.17$  &  $1.27\pm0.20$  &  $0.37\pm0.10$ \\
PT256G5/3    &  $256^3$    &  $5/3$  &  $13.4\pm0.7$  &  $6.3\pm0.3$  &  $2.15\pm0.12$  &  $1.27\pm0.04$  &  $1.98\pm0.11$  &  $1.30\pm0.13$  &  $0.32\pm0.06$ \\
\hline
\end{tabular}\\\vspace{0.05cm}
\raggedright{
\textbf{Notes.} Column 1: simulation name. Columns 2, 3: grid resolution and polytropic exponent $\Gamma$ in Equation~(\ref{eq:eos}). Columns 4, 5: volume-weighted and mass-weighted rms Mach number. Columns 6, 7: volume-weighted and mass-weighted standard deviation of logarithmic density fluctuations in the simulations (Sim $\sigsv$ and Sim $\sigsm$). Columns 8, 9: same as columns 6 and 7, but fitted via the \citet{Hopkins2013PDF} PDF, Equation~(\ref{eq:hopkinspdf}). Column 10: intermittency parameter $\theta$ from the \citet{Hopkins2013PDF} PDF fit. Note also the relations between $\theta$, $\sigsv$ and $\sigsm$, given by Equation~(\ref{eq:sigsvsigsm}).
}
\end{table*}

We use the \textsc{flash} code \citep{FryxellEtAl2000,DubeyEtAl2008}, version~4, to solve the compressible hydrodynamical equations on 3D, uniform, periodic grids of fixed side length $L$. To guarantee stability and accuracy of the numerical solution, we use the HLL5R positive-definite Riemann solver \citep{WaaganFederrathKlingenberg2011}.

We drive turbulence by applying a stochastic acceleration field ${\bf F_\mathrm{stir}}$ as a momentum and energy source term. ${\bf F_\mathrm{stir}}$ only contains large-scale modes, $1<\left|\mathbf{k}\right|L/2\pi<3$, where most of the power is injected at the $\kinj=2$ mode in Fourier space, i.e., on half of the box size (for simplicity, we drop the wavenumber unit $L/2\pi$ in the following). This large-scale driving is favoured by molecular cloud observations \citep[e.g.,][]{OssenkopfMacLow2002,HeyerWilliamsBrunt2006,BruntHeyerMacLow2009,RomanDuvalEtAl2011}. The turbulence on smaller scales, $k\geq3$, is not directly affected by the driving and develops self-consistently. We use the stochastic Ornstein-Uhlenbeck process to generate ${\bf F_\mathrm{stir}}$ with a finite autocorrelation timescale \citep{EswaranPope1988,SchmidtHillebrandtNiemeyer2006}, set to the turbulent crossing time on the largest scales of the system, $T\equiv L/(2\sigma_v)$, where $\sigma_v$ is the velocity dispersion on the integral scale, $L/2$ \citep[for details, see][]{SchmidtEtAl2009,FederrathDuvalKlessenSchmidtMacLow2010,KonstandinEtAl2012}. The turbulent driving used here excites a natural mixture of solenoidal and compressible modes, corresponding to a turbulent driving parameter $b=0.4$ \citep{FederrathDuvalKlessenSchmidtMacLow2010}. All simulations were run for 10 turbulent crossing times, $10\,T$, allowing us to study convergence in time and to average PDFs and spectra in the regime of fully developed turbulence.

The setup and numerical methods are the same as in previous studies \citep[e.g.,][]{FederrathKlessenSchmidt2008,FederrathKlessenSchmidt2009,FederrathDuvalKlessenSchmidtMacLow2010,FederrathEtAl2011PRL,FederrathSchoberBovinoSchleicher2014,Federrath2013}. The important difference is that instead of an isothermal EOS, we here use a polytropic EOS,
\begin{equation} \label{eq:eos}
P = P_0 \left(\rho/\rho_0\right)^\Gamma,
\end{equation}
with the mean density $\rho_0$ and the normalisation pressure $P_0=\rho_0\cs^2(\Gamma\!=\!1)$, i.e., normalised with respect to the isothermal sound speed ($\Gamma=1$). As these simulations are scale free, we set $L=1$, $\rho_0=1$, $P_0=1$, and provide measurements of the gas density, temperature, pressure, and sound speed always relative to the respective mean values, which covers all the physics in the simulations and allows us to scale the simulation quantities to any arbitrary cloud size for comparisons with observations. In order to study the influence of temperature variations on the statistics of supersonic non-isothermal turbulence (in particular its effect on the density PDF and SFR), we vary the polytropic exponent $\Gamma$ in Equation~(\ref{eq:eos}) and perform simulations with
\begin{itemize}
\item $\Gamma=0.7$\quad(soft EOS), \item $\Gamma=1$\quad(isothermal EOS), \item $\Gamma=5/3$\quad(stiff EOS).
\end{itemize}

In order to test and establish numerical convergence, we run simulations with grid resolutions of $256^3$, $512^3$, $1024^3$, and $2048^3$ compute cells. Table~\ref{tab:sims} lists the key parameters of all our polytropic turbulence simulations.

\section{The filamentary structure of polytropic turbulence} \label{sec:structure}

\begin{figure*}
\centerline{\includegraphics[width=0.97\linewidth]{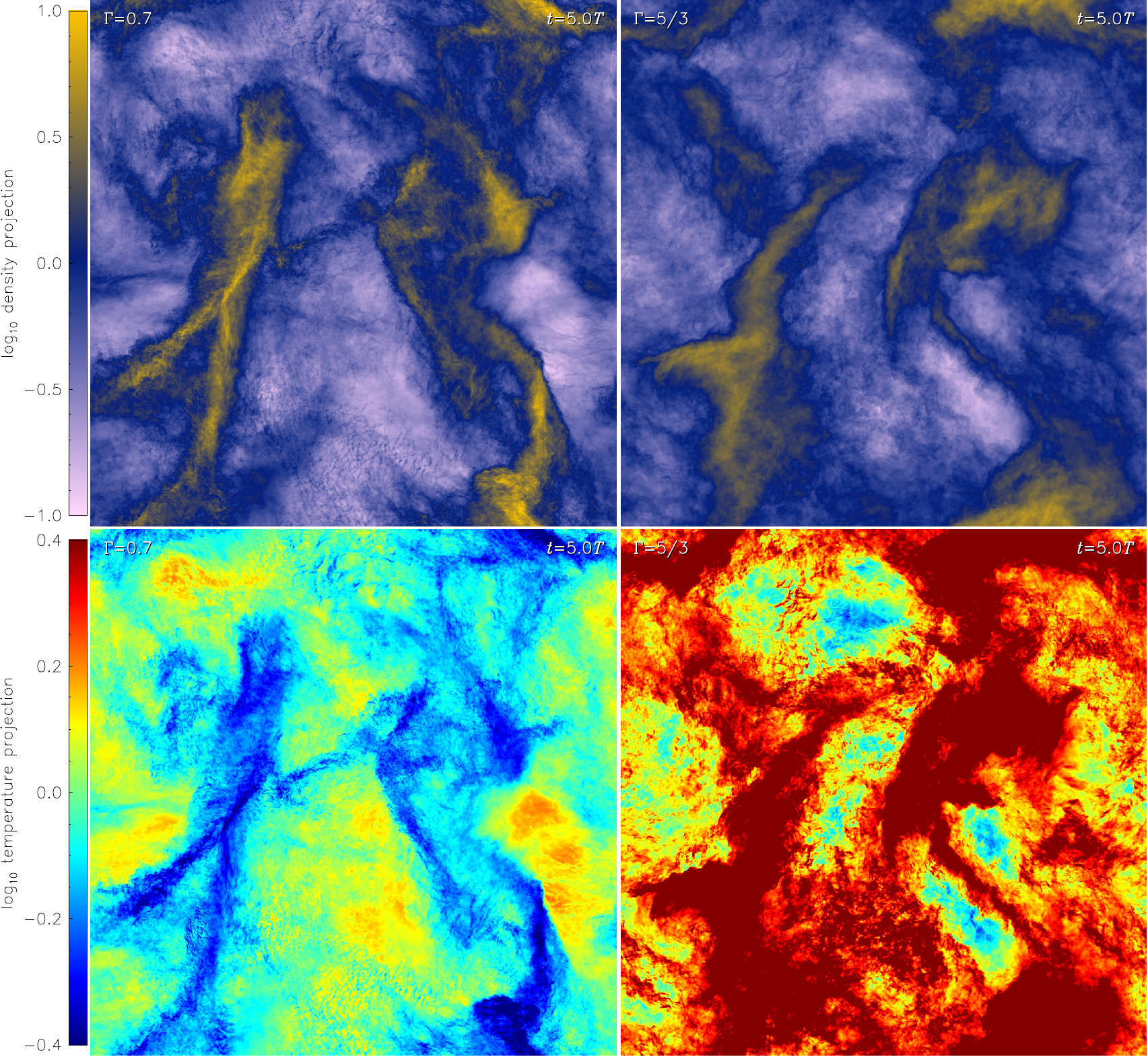}}
\caption{Density projections (top) and temperature projections (bottom) for our non-isothermal turbulence simulations with polytropic exponent $\Gamma=0.7$ (left) and $\Gamma=5/3$ (right) when the turbulence is fully developed. We clearly see how gas with $\Gamma<1$ cools when it is compressed in dense shocks, while gas with $\Gamma > 1$ heats up during compression. Real molecular clouds are in the \mbox{$\Gamma\lesssim1$} regime over a wide range of gas densities and only when dense cores form do they turn into the $\Gamma>1$ regime as the gas becomes optically thick. We also see that lower $\Gamma$ results in a more fragmented density cloud on small scales, while $\Gamma>1$ smoothes out density fluctuations. These polytropic turbulence simulations use unprecedented resolutions of $2048^3$ grid cells. The units are always shown normalised to the respective average values. An animation of this still frame is available in the online version of the journal.}
\label{fig:imagesproj}
\end{figure*}

Figure~\ref{fig:imagesproj} shows projections (integration along the line of sight) of the 3D density (top panels) and temperature (bottom panels) in our numerical simulations with $\Gamma=0.7$ (left-hand panels) and $\Gamma=5/3$ (right-hand panels). These column-density and column-temperature projections are close to what an observer would see in a real molecular cloud observation. Several large- and small-scale filaments are readily identifiable and would deserve deeper analyses along the lines of \citet{AndreEtAl2010,HenningEtAl2010,MenshchikovEtAl2010,SchmalzlEtAl2010,ArzoumanianEtAl2011,HillEtAl2011,HennemannEtAl2012,PerettoEtAl2012,SchneiderEtAl2012,HacarEtAl2013}. We clearly see that the small-scale column density structure strongly depends on polytropic $\Gamma$ with $\Gamma<1$ leading to a more fragmented density field with filaments of high density contrast, while $\Gamma>1$ smoothes out small-scale density contrasts. Comparing these different $\Gamma$ models with observations of filaments could reveal important clues about the thermodynamic state of molecular clouds.

The temperature structure shown in the bottom panels of Figure~\ref{fig:imagesproj} follows our expectations, such that for $\Gamma=0.7$, the dense gas clouds in the simulation are colder than their surrounding low-density gas, while the opposite applies in the $\Gamma=5/3$ case. Real molecular clouds are in the \mbox{$\Gamma\lesssim1$} regime for number densities of $n\sim10^2$--$10^5\,\cm^{-3}$ \citep{OmukaiEtAl2005,GloverMacLow2007a,GloverMacLow2007b,GloverFederrathMacLowKlessen2010}. The gas becomes optically thick and turns into the $\Gamma>1$ regime only in the very dense cores for $n\gtrsim10^{10}\,\cm^{-3}$ \citep[][]{OmukaiEtAl2005,MasunagaInutsuka2000}. We note that our numerical experiments are scale free and can thus be applied to both cases, depending on $\Gamma$.

\begin{figure*}
\centerline{\includegraphics[width=0.97\linewidth]{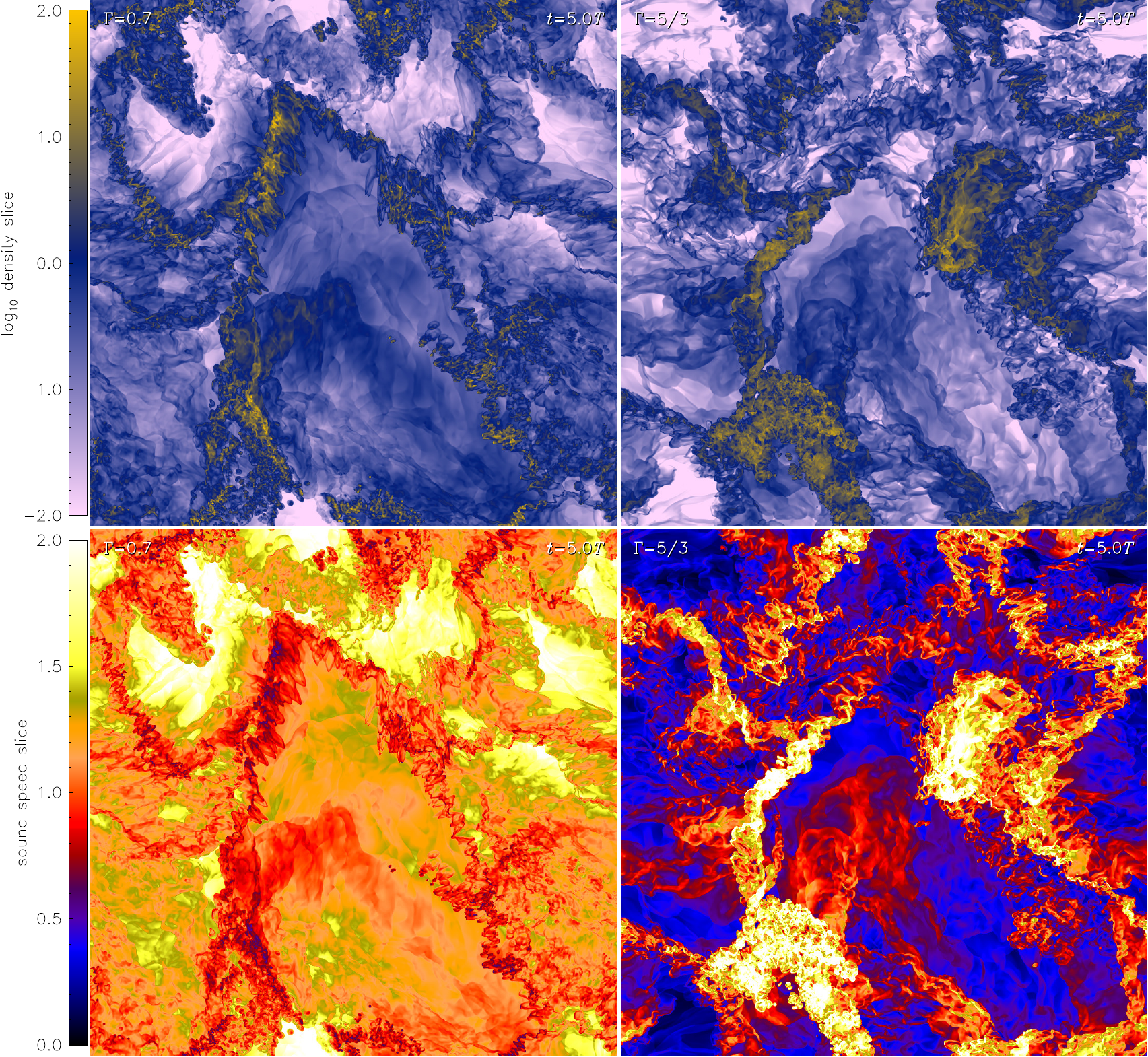}}
\caption{As Figure~\ref{fig:imagesproj}, but here we show slices of the gas density (top) and slices of the sound speed through the mid-plane of our 3D turbulent domain with $2048^3$ grid cells. For polytropic $\Gamma<1$ (left-hand panels), the sound speed decreases in the shocks, while it increases for $\Gamma>1$ (right-hand panels), which has important consequences for the Mach number and density PDFs, and for the SFR. An animation of this still frame is available in the online version of the journal.}
\label{fig:imagesslice}
\end{figure*}

Figure~\ref{fig:imagesslice} shows slices through the mid-plane of our computational domain. The top and bottom panels display density and sound speed, respectively. These spatial slices reveal individual shocks and multi-shock interactions. As expected, the sound speed drops in the shocks if $\Gamma<1$, while it increases in the shocks if $\Gamma>1$. We also see that most of the volume is occupied by post-shock gas, i.e., gas that is currently in a state of expansion or rarefaction, which exhibits the opposite behaviour to the compressed gas: rarefied gas heats up for $\Gamma<1$ and cools down for $\Gamma>1$. Thus, we expect the volume-weighted average sound speed to be higher for $\Gamma=0.7$ than for $\Gamma=5/3$. In contrast, the mass-weighted average sound speed (which is dominated by shocked regions) is expected to be lower for $\Gamma=0.7$ than for $\Gamma=5/3$.

\section{The time evolution of polytropic turbulence} \label{sec:tevol}

\begin{figure*}
\centerline{\includegraphics[width=0.97\linewidth]{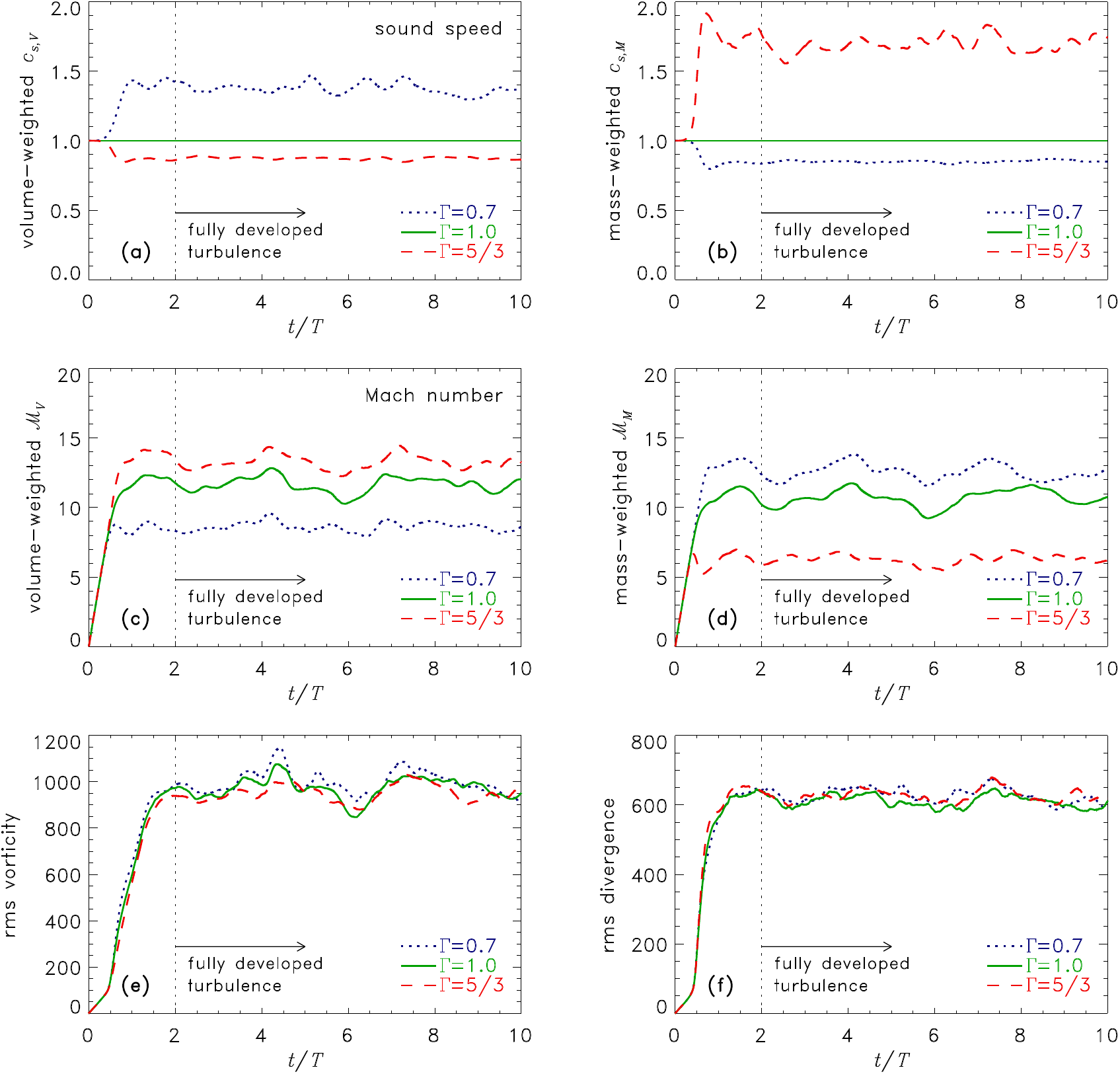}}
\caption{Time evolution of the volume-weighted average sound speed $\csv$ (panel a), the mass-weighted average sound speed $\csm$ (panel b), the volume-weighted root-mean-square (rms) Mach number $\machv$ (panel c), the mass-weighted rms Mach number $\machm$ (panel d), the rms vorticity $\langle(\nabla\times\vect{v})^2\rangle^{1/2}$ (panel e), and the rms divergence $\langle(\nabla\cdot\vect{v})^2\rangle^{1/2}$ (panel f) for simulations with polytropic exponent $\Gamma=0.7$ (dotted), $\Gamma=1.0$ (solid), and $\Gamma=5/3$ (dashed). The turbulence is fully developed after two turbulent crossing times, $t\geq2\,T$, as indicated by the vertical dotted lines in each panel.}
\label{fig:tevol}
\end{figure*}

We start all our simulations with gas initially at rest and with a homogeneous density $\rho_0$ in a 3D periodic box. The turbulent driving accelerates the gas to our target velocity dispersion $\sigma_v$. Depending on the choice of polytropic exponent ($\Gamma=0.7$, $1$, $5/3$) we expect different average sound speeds and Mach numbers. Figure~\ref{fig:tevol} shows the time evolution of the average sound speed (top panels) and rms Mach number (middle panels). The left-hand panels show the volume-weighted averages and the right-hand panels show the mass-weighted averages. After two turbulent crossing times, $t\geq2\,T$, all volume- and mass-weighted averages are converged in time and only fluctuate by about 10\% around their typical, time-averaged value.

The top left-hand panel of Figure~\ref{fig:tevol} shows that the volume-weighted sound speed increases for $\Gamma<1$, while it decreases for $\Gamma>1$ (we always use the well-studied isothermal case, $\Gamma=1$, as a reference). The mass-weighted averages shown in the top right-hand panel, on the other hand, exhibit the opposite behaviour. This can be readily understood when we consider how the sound speed changes upon compression of the gas in a shock for different $\Gamma$. For example, if $\Gamma<1$, then a compression leads to cooling, decreasing the sound speed in the shocks. In contrast, the gas heats up in expanding, rarefied regions. Since most of the volume is always in a state of rarefaction or expansion, because the shocks only occupy a small fraction of the volume, the volume-averaged sound speed increases for $\Gamma=0.7$ compared to $\Gamma=1$. On the other hand, since most of the mass is in shocks and the gas cools when it is compressed for $\Gamma<1$, the mass-weighted sound speed decreases for $\Gamma=0.7$ compared to $\Gamma=1$. The opposite happens for gas with $\Gamma>1$, which heats up during a compression and cools down during a rarefaction: $\csv$ decreases, while $\csm$ increases with respect to the isothermal case.

The middle panels of Figure~\ref{fig:tevol} show the volume-weighted Mach number $\machv=\sigma_v/\csv$ (left-hand panel), and the mass-weighted rms Mach number $\machm=\sigma_v/\csm$, respectively. Since the velocity dispersion $\sigma_v$ is the same in all our numerical experiments, the dependence of the Mach number on $\Gamma$ is basically the inverse of the dependence of the sound speed on $\Gamma$. It is worth pointing out that the volume- and mass-weighted rms Mach numbers for the isothermal case ($\Gamma=1$) are very similar\footnote{The mass-weighted rms Mach number for $\Gamma=1$ is $\sim9\%$ smaller than the volume-weighted rms (see Table~\ref{tab:sims}), because the mass-weighted rms puts more weight on the shocks, which have somewhat smaller velocity dispersion, because they represent stagnation points of the overall turbulent flow.}, because the sound speed is the same, while it is important to distinguish volume- and mass-weighted averages for non-isothermal gases ($\Gamma\neq1$).

Finally, the bottom panels of Figure~\ref{fig:tevol} show the rms vorticity ($\nabla\times\vect{v}$) and rms divergence ($\nabla\cdot\vect{v}$) of the turbulent velocity field, which are largely insensitive to changes in $\Gamma$. This is because the turbulent driving primarily determines the amount of solenoidal and compressible modes in the velocity field, while baroclinic vorticity production \citep{MeeBrandenburg2006,DelSordoBrandenburg2011,FederrathEtAl2011PRL}, which is only possible in non-isothermal gases such as our $\Gamma\ne1$ cases, is negligible compared to the directly induced solenoidal and compressible modes.

\section{The density PDF of polytropic turbulence} \label{sec:pdfs}

The density PDF provides an important statistical measure of the distribution of gas densities in the interstellar medium of galaxies \citep[][]{BerkhuijsenFletcher2008,HughesEtAl2013} and in molecular clouds in the Milky Way. The PDF has recently attracted attention, because it provides us with valuable information about the potential of a cloud to form stars \citep[][and references therein]{FederrathKlessen2012,PadoanEtAl2014}. The key feature of the PDF is that we can use it to determine the dense gas mass fraction of a cloud, capable of forming stars. Submillimetre observations of the column-density PDF in the spiral arms of the Milky Way show that narrower PDFs with less dense gas are typically found in rather quiescent clouds (in terms of star formation), while clouds with a wider PDF and correspondingly higher dense gas mass fraction are actively forming stars \citep{KainulainenEtAl2009,SchneiderEtAl2012,SchneiderEtAl2013,GinsburgFederrathDarling2013,KainulainenFederrathHenning2013,KainulainenFederrathHenning2014}. Recent ALMA observations in the Galactic Centre by \citet{RathborneEtAl2014} also reveal the typical features seen in the density PDFs produced by supersonic turbulence \citep[e.g.,][]{FederrathKlessenSchmidt2008,FederrathKlessen2013}.

\subsection{The density PDF in isothermal gas}

Most of the underlying theoretical and numerical work on the density PDF is based on the assumption of isothermal gas ($\Gamma=1$) \citep{PadoanNordlundJones1997,Klessen2000,KritsukEtAl2007,LemasterStone2008,FederrathDuvalKlessenSchmidtMacLow2010,BruntFederrathPrice2010a,BruntFederrathPrice2010b,PriceFederrathBrunt2011,KonstandinEtAl2012ApJ,MolinaEtAl2012,MicicEtAl2012,MoraghanKimYoon2013,Federrath2013,LeeChangMurray2014}, leading to a lognormal PDF in the logarithmic density contrast $s\equiv\ln(\rho/\rho_0)$,
\begin{equation} \label{eq:lognormalpdf}
p_V(s)=\frac{1}{\left(2\pi\sigsv^2\right)^{1/2}}\exp\left(-\frac{(s-s_0)^2}{2\sigsv^2}\right).
\end{equation}
The lognormal PDF contains two parameters: 1) the volume-weighted density variance $\sigsv$ and 2) the mean value $s_0$, which is related to the variance by $s_0=-\sigsv^2/2$ due to mass conservation \citep{Vazquez1994,FederrathKlessenSchmidt2008}. There are a few important studies where the influence of temperature variations on the PDF has been explored \citep{Vazquez1994,PassotVazquez1998,WadaNorman2001,KritsukNorman2002,LiKlessenMacLow2003,AuditHennebelle2005,AuditHennebelle2010,HennebelleAudit2007,KissmannEtAl2008,SeifriedEtAl2011,GazolKim2013}, showing that the PDF tends to depart significantly from the lognormal form given by Equation~(\ref{eq:lognormalpdf}) if the gas is non-isothermal. Here we extend these previous studies to much higher resolution in order to determine the density PDF of highly supersonic, non-isothermal, polytropic turbulence and to provide a theoretical model for the width of the PDF and for the SFR as a function of the polytropic exponent $\Gamma$.

\subsection{The density PDF in non-isothermal gas}

\begin{figure*}
\centerline{\includegraphics[width=0.97\linewidth]{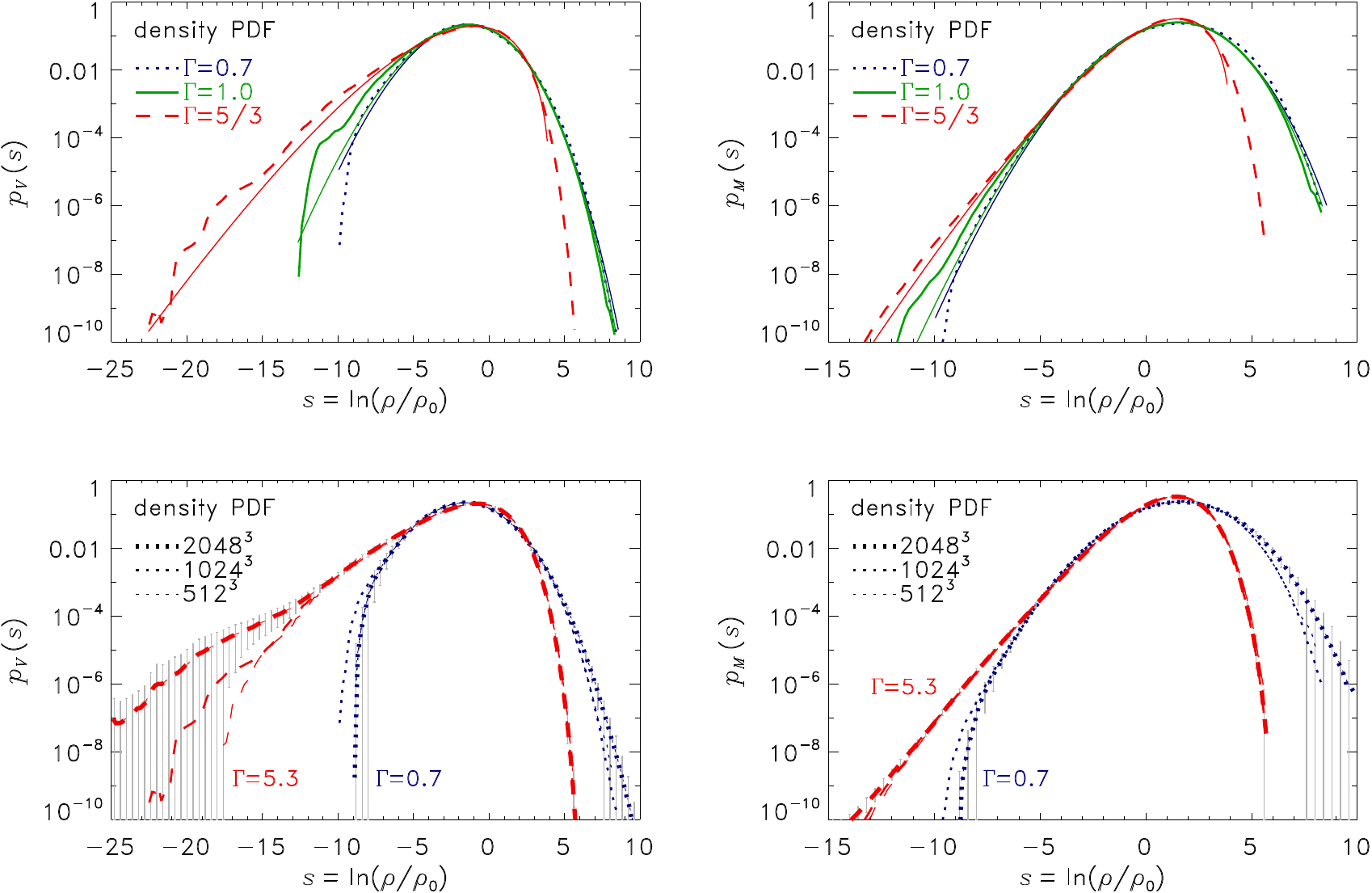}}
\caption{Top panels: the density PDF of supersonic, polytropic turbulence with $\Gamma=0.7$ (dotted), $\Gamma=1$ (solid), and $\Gamma=5/3$ (dashed). The left panel shows the volume-weighted PDFs, while the right panel shows the mass-weighted PDFs of the logarithmic density contrast $s\equiv\ln(\rho/\rho_0)$. Each simulation PDF is fitted with the \citet{Hopkins2013PDF} intermittency PDF model, Equation~(\ref{eq:hopkinspdf}), shown as thin solid lines. They provide excellent fits to the PDFs of isothermal ($\Gamma=1$) and non-isothermal ($\Gamma\neq1$) intermittent turbulence with two fit parameters (the standard deviation $\sigs$ and the intermittency parameter $\theta$), listed in the last three columns of Table~\ref{tab:sims}. Bottom panels: same as top panels, but showing a resolution study with $512^3$, $1024^3$, and $2048^3$ grid cells. The 1-sigma time variations are shown as error bars only for the $N_\mathrm{res}^3=2048^3$ models, but the time variations are similar for all resolutions.}
\label{fig:pdfs}
\end{figure*}

Figure~\ref{fig:pdfs} shows the density PDFs of $s\equiv\ln(\rho/\rho_0)$ obtained in our simulations with polytropic $\Gamma=0.7$, $1$, and $5/3$. We immediately see that the density PDF depends on $\Gamma$. For $\Gamma>1$, we find that a power-law tail develops at low densities, consistent with earlier numerical work in one-dimensional geometry \citep{PassotVazquez1998}. For $\Gamma=0.7$ and $\Gamma=1$, we see a more symmetric distribution, but we do not see a clear power-law tail at high densities for $\Gamma<1$. This is because our simulations are in 3D and multiple shock interactions lead to a more lognormal distribution as a result of the central limit theorem \citep{Vazquez1994} than in the one-dimensional simulations by \citet{PassotVazquez1998}. 

We apply fits to all PDFs in the top panels of Figure~\ref{fig:pdfs} for different $\Gamma$, shown as thin solid lines. The fit function is given by the \citet{Hopkins2013PDF} intermittency PDF model,
\begin{align} \label{eq:hopkinspdf}
p_V(s) = I_1\left(2\sqrt{\lambda\,\omega(s)}\right)\exp\left[-\left(\lambda+\omega(s)\right)\right]\sqrt{\frac{\lambda}{\theta^2\,\omega(s)}}\,,\nonumber\\
\lambda\equiv\sigsv^2/(2 \theta^2),\quad\omega(s)\equiv\lambda/(1+ \theta)-s/ \theta \;\; (\omega\geq0),
\end{align}
where $I_1(x)$ is the modified Bessel function of the first kind. Equation~(\ref{eq:hopkinspdf}) is motivated and explained in detail in \citet{Hopkins2013PDF}. It contains two parameters: 1) the volume-weighted standard deviation of logarithmic density fluctuations $\sigsv$, and 2) the intermittency parameter $ \theta$. The volume-weighted and the mass-weighted variances are given and related by \citep{Hopkins2013PDF}
\begin{equation} \label{eq:sigsvsigsm}
\sigsv^2 = 2\lambda\theta^2 = \sigsm^2 (1+\theta)^3.
\end{equation}
In the zero-intermittency limit ($\theta\to0$), Equation~(\ref{eq:hopkinspdf}) simplifies to the lognormal PDF, Equation~(\ref{eq:lognormalpdf}). \citet{Hopkins2013PDF} shows that the intermittency form of the PDF (Equation~\ref{eq:hopkinspdf}) provides excellent fits to density PDFs from turbulence simulations with extremely different properties (solenoidal, mixed, and compressive driving, Mach numbers from 0.1 to 20, and varying magnetic field strengths). It has also been used to study convergence of the PDF with numerical resolution \citep{Federrath2013}. Here we show in Figure~\ref{fig:pdfs} that Equation~(\ref{eq:hopkinspdf}) furthermore provides very good fits to the density PDFs from simulations with different polytropic exponent $\Gamma$. The PDF fit parameters are listed in the last three columns of Table~\ref{tab:sims}.

The bottom panels of Figure~\ref{fig:pdfs} show a resolution study, comparing the density PDFs obtained for grid resolutions of $512^3$, $1024^3$, and $2048^3$ compute cells. We find numerical convergence over a wide range of densities around the peak of all distributions. The highly intermittent low-density tail for $\Gamma=5/3$ gas shows strong variations in the volume-weighted form of the PDF (left-hand panel), but is nearly converged in the mass-weighted representation (right-hand panel). We find that higher resolution is required to obtain numerical convergence if $\Gamma<1$, because the gas is much more fragmented and filamentary on small scales than for $\Gamma>1$, as we have seen in the projections of Figure~\ref{fig:imagesproj}. However, the variations with resolution are of the same order or smaller than the 1-sigma temporal variations shown as grey error bars for the $2048^3$ simulations in Figure~\ref{fig:pdfs}. Table~\ref{tab:sims} lists all PDF properties for each simulation model and resolution, demonstrating convergence of the main PDF properties, $\sigsv$, $\sigsm$, and $\theta$.

\subsection{The column-density PDF in polytropic gas}

The column-density PDF has become an important statistical measure for the density structure of the interstellar medium in external galaxies \citep{BerkhuijsenFletcher2008,HughesEtAl2013}, in molecular clouds in the Milky Way \citep{KainulainenEtAl2009,SchneiderEtAl2012,SchneiderEtAl2013,GinsburgFederrathDarling2013,KainulainenFederrathHenning2013,KainulainenFederrathHenning2014}, and in the Galactic central molecular zone \citep{RathborneEtAl2014}. Here we produce column-density PDFs from our polytropic turbulence simulations for comparison with observations.

\begin{figure*}
\centerline{\includegraphics[width=0.99\linewidth]{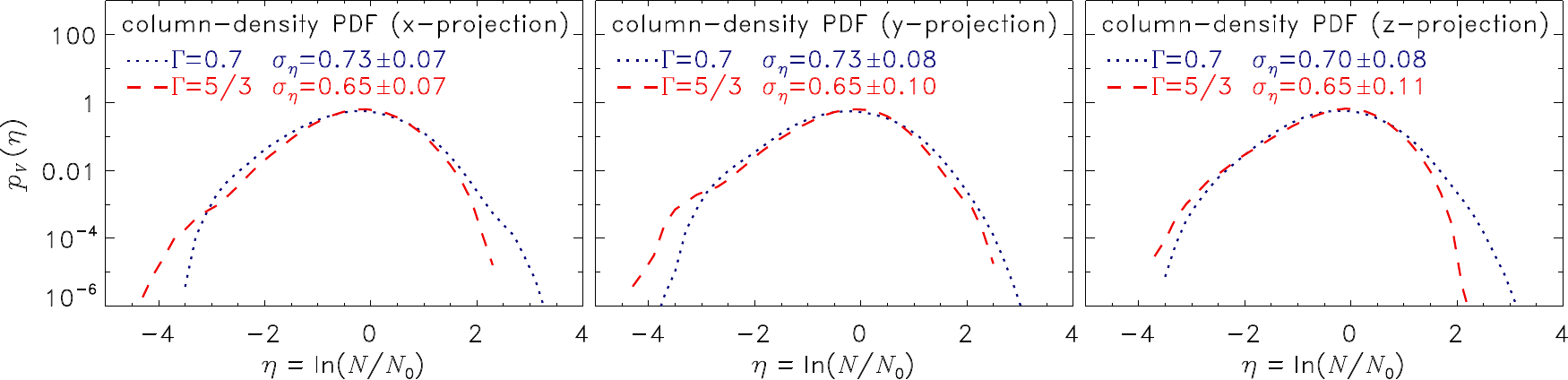}}
\caption{Column-density PDFs of the logarithmic column density contrast $\eta\equiv\ln(N/N_0)$ in the simulations with polytropic $\Gamma=0.7$ (dotted) and $\Gamma=5/3$ (dashed). The panels show the column-density PDFs for projections along the $x$-axis (left-hand panel), $y$-axis (middle panel), and $z$-axis (right-hand panel). The standard deviation $\sigma_\eta$ is given in each panel and is consistent with observations. For $\Gamma>1$, the PDFs are slightly skewed towards lower column densities, while they roughly follow lognormal distributions for $\Gamma\lesssim1$, consistent with the trend seen in the volumetric PDFs (cf.~Figure~\ref{fig:pdfs}).}
\label{fig:pdfs_coldens}
\end{figure*}

Figure~\ref{fig:pdfs_coldens} shows the column density PDFs in our simulations with $\Gamma=0.7$ and $\Gamma=5/3$. Each panel represents a different line-of-sight projection, along the $x$-axis (left-hand panel), the $y$-axis (middle panel), and the $z$-axis (right-hand panel). We find that the column-density PDFs for $\Gamma\lesssim1$ are close to lognormal distributions, while the PDFs for $\Gamma>1$ develop a power-law tail towards low column densities. This trend with $\Gamma$ is consistent with what we found for the volumetric density PDFs in Figure~\ref{fig:pdfs}. The differences with $\Gamma$ are less pronounced in the column-density PDFs, because the projection averages out features seen only in the 3D distributions. Nevertheless, the systematic differences between column-density PDFs from gas with different $\Gamma$ are significant and recovered in all of the three line-of-sight projections.

The standard deviation of the column-density contrast, $\sigma_\eta$, is given in each panel of Figure~\ref{fig:pdfs_coldens}. Consistent with the volumetric PDFs, we find that $\sigma_\eta$ decreases slightly (but systematically) with increasing $\Gamma$. Since $\sigma_\eta$ is the column-density contrast normalised to the mean column density $N_0$, it can be easily and directly compared with observations. \cite{SchneiderEtAl2012} find $\sigma_\eta\sim0.63$ in Herschel observations of the Rosette molecular cloud, consistent with, but slightly lower than our numerical simulations. However, the Mach number in Rosette is $\mach\sim7$, while here we have $\mach\sim10$, so we expect a somewhat lower $\sigma_\eta$ in the observations. Furthermore, we have not included magnetic fields in these simulations, which would further reduce the column-density variance \citep{MolinaEtAl2012}. Given these limitations, the agreement between the simulations and the observations is encouraging.

\subsection{The pressure, sound speed, and Mach number distributions of polytropic turbulence}

\begin{figure*}
\centerline{\includegraphics[width=0.97\linewidth]{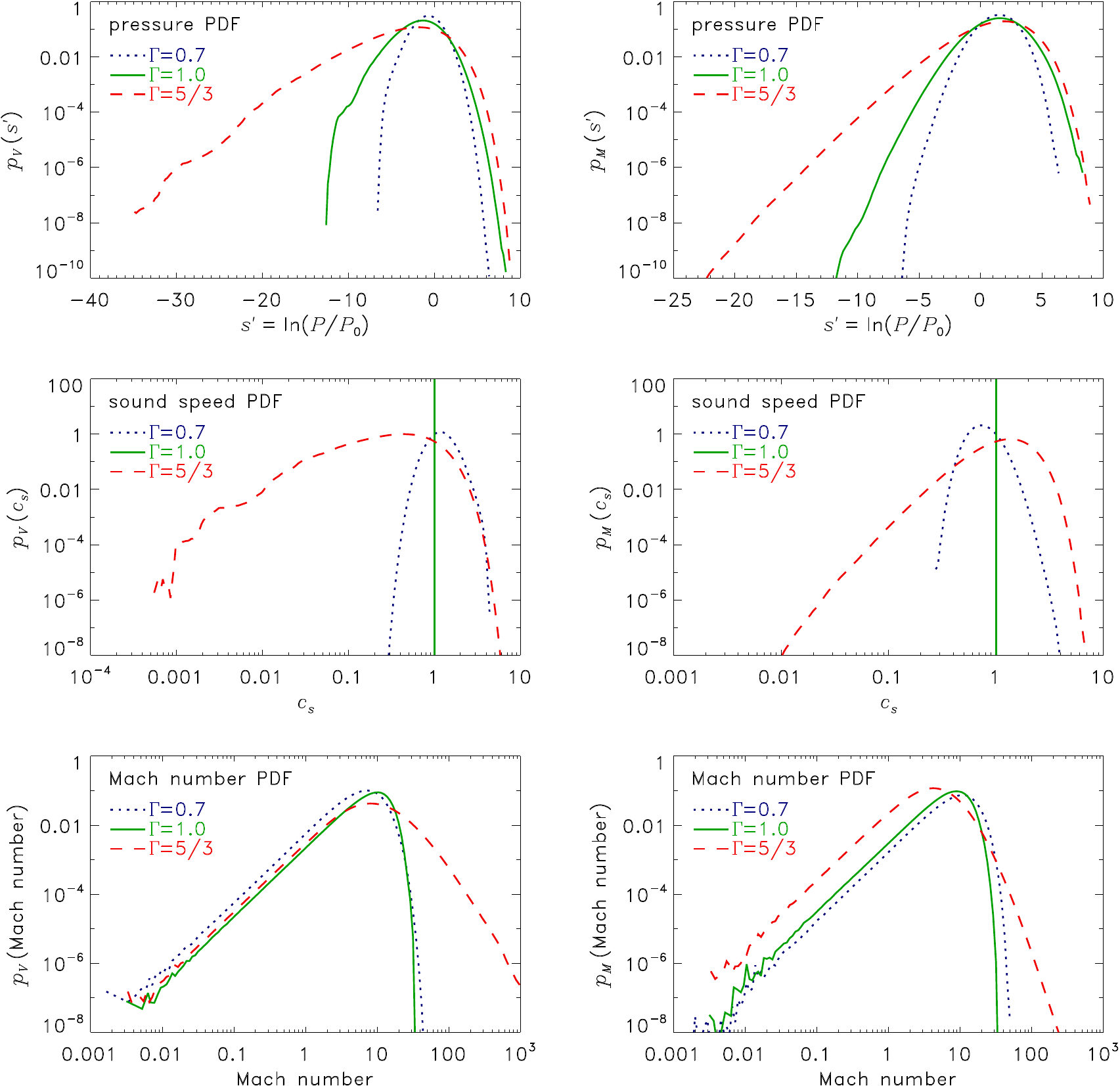}}
\caption{Pressure PDF (top), sound speed PDF (middle), and Mach number PDF (bottom) for simulations with polytropic $\Gamma=0.7$, $1.0$, and $5/3$. The left-hand panels show the volume-weighted form of the PDF and the right-hand panels show the mass-weighted form. As expected, $\Gamma>1$ leads to a wider pressure distribution than $\Gamma<1$. The sound speed PDF for $\Gamma=1$ is a delta function, because the sound speed is constant for $\Gamma=1$. The Mach number PDFs exhibit very wide distributions and they always have power-law tails towards small Mach numbers. The local Mach number can reach several hundreds to thousands for $\Gamma=5/3$.}
\label{fig:pdfsmisc}
\end{figure*}

Figure~\ref{fig:pdfsmisc} shows the pressure PDFs (top panels), sound speed PDFs (middle panels), and Mach number PDFs (bottom panels). As expected for a polytropic EOS, the pressure distribution for $\Gamma>1$ is significantly wider than for $\Gamma<1$. The sound speed distributions follow a similar trend. They tend to have a wide tail towards low sound speeds for $\Gamma>1$, whereas they are narrow and close to lognormal distributions for $\Gamma<1$. The sound speed is constant in isothermal gas ($\Gamma=1$), so the PDF is a delta function. We find a wide distribution of Mach numbers with power-law tails towards small Mach numbers in all cases. The Mach number PDF for $\Gamma=5/3$ also develops a power-law tail at high Mach numbers, where the Mach numbers can locally reach extremely high values of several hundred to a few thousand.

\begin{figure}
\centerline{\includegraphics[width=0.97\linewidth]{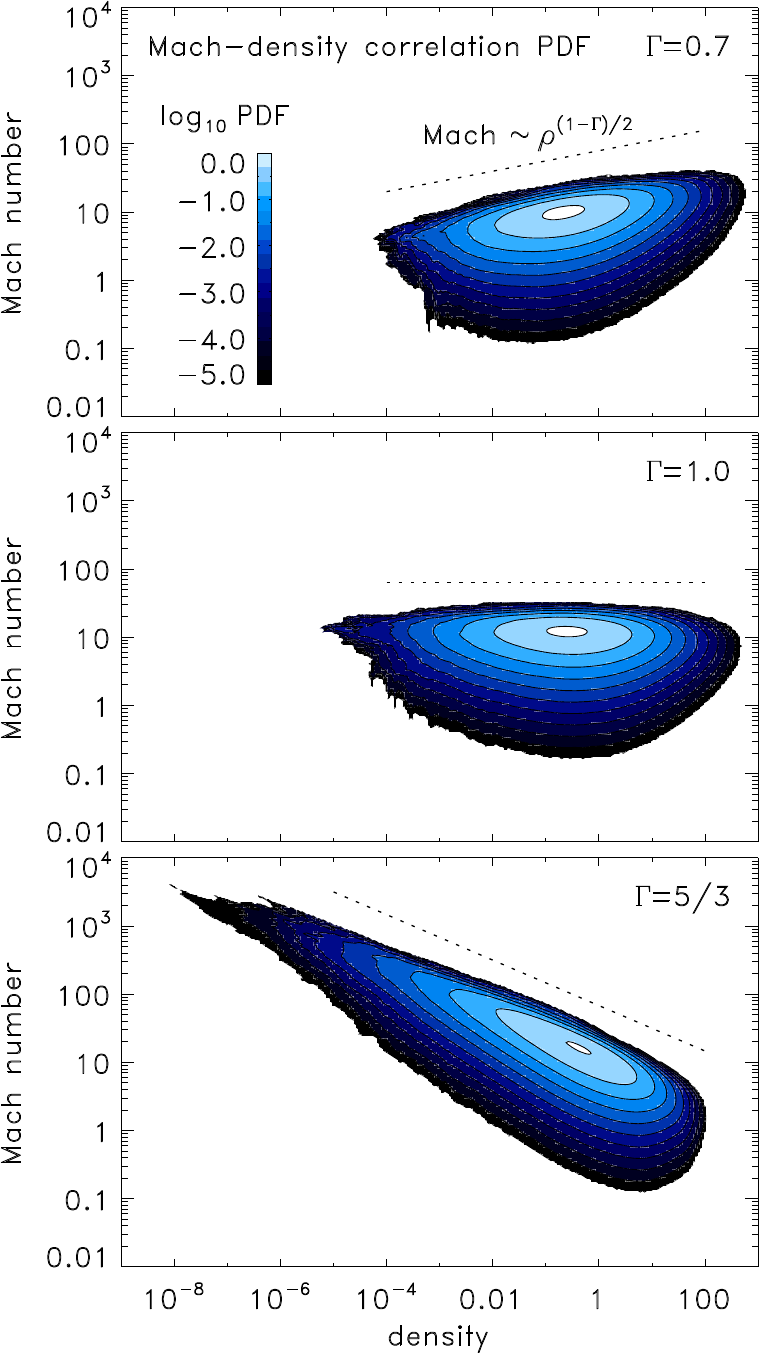}}
\caption{Density -- Mach number correlation PDFs for $\Gamma=0.7$ (top), $\Gamma=1$ (middle), and $\Gamma=5/3$ (bottom). For a polytropic EOS, the theoretical expectation for the average run of the Mach number as a function of density is given by the dotted lines in each panel, $\mach\sim\rho^{(1-\Gamma)/2}$, which matches the simulation outcome. For $\Gamma=1$, we expect no net correlation between local Mach number and local density. For $\Gamma=5/3$, we see that the highest Mach numbers of a few thousand are reached in very low-density gas, which is a result of the adiabatic cooling of expanding, low-density gas, drastically reducing the local sound speed.}
\label{fig:2dpdfs}
\end{figure}

In preparation for our derivation of the density variance -- Mach number relation for polytropic turbulence, presented in the next section, we show here density -- Mach number correlation PDFs in Figure~\ref{fig:2dpdfs}. These have been used earlier to explain the nearly lognormal form of the density PDF in isothermal gas \citep{PassotVazquez1998,KritsukEtAl2007,AuditHennebelle2010,FederrathDuvalKlessenSchmidtMacLow2010}. If the gas is isothermal, there should not be any net correlation between density and Mach number, which is seen in the middle panel of Figure~\ref{fig:2dpdfs}. In contrast, for $\Gamma\neq1$, it is straightforward to show that the local Mach number must depend on the local density, following $\mach\sim\cs^{-1}\sim(P/\rho)^{-1/2}\sim\rho^{1/2}\rho^{-\Gamma/2}\sim\rho^{(1-\Gamma)/2}$. This is indeed the case, as seen in the top and bottom panels of Figure~\ref{fig:2dpdfs}, showing our simulation results for $\Gamma=0.7$ and $\Gamma=5/3$, respectively. The dotted line is the theoretical prediction for the average run of the Mach number as a function of gas density, $\mach\sim\rho^{(1-\Gamma)/2}$, which provides an excellent match to our simulations.

\section{Density variance -- Mach number relation in polytropic gases} \label{sec:densityvariancemachrelation}

The density variance -- Mach number relation is a key ingredient to theoretical models of the SFR \citep{KrumholzMcKee2005,PadoanNordlund2011,HennebelleChabrier2011,FederrathKlessen2012}, the star formation efficiency \citep{Elmegreen2008}, the IMF of stars \citep{HennebelleChabrier2008,HennebelleChabrier2009,HennebelleChabrier2013,Hopkins2013IMF,ChabrierHennebelle2011,ChabrierEtAl2014}, and the Kennicutt-Schmidt relation \citep{Federrath2013sflaw}.

The $\sigs$--$\mach$ relation has been studied numerically for isothermal gas by \citet{PadoanNordlund2011}, \citet{PriceFederrathBrunt2011}, \citet{KonstandinEtAl2012ApJ}, \citet{Seon2012}, and \citet{MolinaEtAl2012}, resulting in
\begin{equation}
\sigs^2 = \ln\left(1+b^2\mach^2\frac{\beta}{\beta+1}\right).
\end{equation}
This equation provides us with the density variance as a function of the Mach number $\mach$, the turbulent driving parameter $1/3\leq b \leq1$ \citep{FederrathKlessenSchmidt2008,FederrathDuvalKlessenSchmidtMacLow2010} and the ratio of thermal to magnetic pressure, plasma $\beta$. The theoretical derivation of this important relation has so far only been done for isothermal gas in \citet{PadoanNordlund2011} and \citet{MolinaEtAl2012}. Here we generalise their analysis to the non-isothermal, polytropic regime of turbulence, consisting of a 3D network of interacting non-isothermal shocks and filaments (cf.~Figures~\ref{fig:imagesproj} and~\ref{fig:imagesslice}).

In order to derive the density variance -- Mach number relation for polytropic gas, we must first determine how the density contrast in a single shock depends on the strength of the shock, i.e., how it depends on the Mach number. Once we have derived the density contrast $\rho/\rho_0$ for a single shock, we average over the whole ensemble of such shocks in a cloud with volume $V$ to obtain the density variance of the cloud,
\begin{equation}
\sigma_{\rho/\rho_0}^2 = \frac{1}{V}\int_V{\left(\frac{\rho}{\rho_0}-1\right)^2}\deriv V,
\end{equation}
which can be approximated with the density contrast itself,
\begin{equation} \label{eq:molina_approx}
\sigma_{\rho/\rho_0}^2 \simeq \frac{\rho}{\rho_0}
\end{equation}
for the relevant case of supersonic turbulence, $\rho\gg\rho_0$ \citep{PadoanNordlund2011,MolinaEtAl2012}. Thus, we only have to find the density contrast $\rho/\rho_0$ produced in a non-isothermal, polytropic shock.

\subsection{Density contrast in non-isothermal shocks}

Starting from the Rankine-Hugoniot shock jump conditions in the frame where the shock is stationary, the one-dimensional Euler equations for mass and momentum conservation can be written as
\begin{align}
\rho_1 v_{\parallel,1} & = \rho_2 v_{\parallel,2} \label{eq:rh1} \\
\rho_1 v_{\parallel,1}^2 + P_1 & = \rho_2 v_{\parallel,2}^2 + P_2 \label{eq:rh2}.
\end{align}
Note that $\rho$, $v$ and $P$ are the density, velocity and thermal pressure. The velocity is always perpendicular to the shock front, i.e., parallel to the flow direction, which we denote with a $\parallel$ subscript. The indices 1 and 2 denote pre-shock and post-shock conditions, respectively.

The pressure in polytropic gas is given by
\begin{equation} \label{eq:p}
P = \cs^2 \rho / \Gamma,
\end{equation}
from the sound speed $\cs^2 = \partial P / \partial\rho=\Gamma P/\rho$ for a polytropic EOS, Equation~(\ref{eq:eos}). Inserting this EOS into the momentum Equation~(\ref{eq:rh2}) yields
\begin{align} 
\rho_1\left( v_{\parallel,1}^2 + \csone^2/\Gamma\right) & = \rho_2\left( v_{\parallel,2}^2 + \cstwo^2/\Gamma\right) \label{eq:intermediate1} \\
                                                                       & = \rho_2\left( v_{\parallel,1}^2 \rho_1^2 / \rho_2^2 + \cstwo^2/\Gamma\right), \label{eq:intermediate2}
\end{align}
where we have eliminated the post-shock velocity $v_{\parallel,2}$ in the second step, by use of mass conservation, Equation~(\ref{eq:rh1}). As a consequence of the polytropic EOS, the post-shock sound speed $\cstwo$ is given by
\begin{equation}
\cstwo^2 = \csone^2 \left(\rho_1/\rho_2\right)^{1-\Gamma},
\end{equation}
which we use to replace $\cstwo^2$ in Equation~(\ref{eq:intermediate2}):
\begin{equation}
\rho_1\left( v_{\parallel,1}^2 + \csone^2/\Gamma\right) = \rho_2\left( v_{\parallel,1}^2 \rho_1^2 / \rho_2^2 + \csone^2 \left(\rho_1/\rho_2\right)^{1-\Gamma}/\Gamma\right).
\end{equation}
In order to simplify this equation, we divide both sides by $\rho_1$ and by $\csone^2$ and multiply by $\Gamma$, which yields
\begin{equation}
\Gamma \frac{v_{\parallel,1}^2}{\csone^2} + 1 = \Gamma \frac{v_{\parallel,1}^2}{\csone^2} \left(\frac{\rho_1}{\rho_2}\right) + \left(\frac{\rho_1}{\rho_2}\right)^{-\Gamma}.
\end{equation}
Now we swap indices such that the pre-shock gas is denoted by the average quantities with subscript 0 and we drop the index for the post-shock gas: 
\begin{equation} \label{eq:intermediate3}
\Gamma \frac{v_{\parallel,0}^2}{\cszero^2} + 1 = \Gamma \frac{v_{\parallel,0}^2}{\cszero^2} \left(\frac{\rho_0}{\rho}\right) + \left(\frac{\rho_0}{\rho}\right)^{-\Gamma}.
\end{equation}
Finally, we identify the pre-shock Mach number perpendicular to the shock front (i.e., parallel to the flow direction), $\mach_\parallel=v_{\parallel,0}/\cszero$. If the pre-shock gas is turbulent, then the compressive velocity component perpendicular to the shock is only a fraction $b$ of the total pre-shock velocity $v_0$, such that $v_{\parallel,0} = b\,v_0$. The parameter $b$ is the compressive-to-solenoidal mode mixture parameter, which is typically in the range $1/3\leq b \leq 1$, depending on whether the turbulence is driven by a solenoidal forcing ($b\sim1/3$) or by a compressive forcing $b\sim1$ \citep{FederrathKlessenSchmidt2008,SchmidtEtAl2009,FederrathDuvalKlessenSchmidtMacLow2010,KonstandinEtAl2012,KonstandinEtAl2012ApJ,Federrath2013}. We can thus replace $v_{\parallel,0}^2/\cszero^2=b^2\mach^2$ in Equation~(\ref{eq:intermediate3}), which yields
\begin{equation} \label{eq:intermediate4}
\Gamma b^2\mach^2 + 1 = \Gamma b^2\mach^2 \left(\frac{\rho_0}{\rho}\right) + \left(\frac{\rho_0}{\rho}\right)^{-\Gamma}.
\end{equation}

Rearranging Equation~(\ref{eq:intermediate4}) and collecting terms for the density contrast $x\equiv\rho/\rho_0$ gives
\begin{equation} \label{eq:nasty}
x^\Gamma + \Gamma b^2\mach^2 (x^{-1}-1) - 1 = 0.
\end{equation}
We must solve this equation for the density contrast $x$, but the equation is transcendental and cannot be solved for a general polytropic exponent $\Gamma$. Thus, we have to consider explicit solutions for specific values of $\Gamma$ for which the general Equation~(\ref{eq:nasty}) can be solved. To this end, we chose to explore solutions for $\Gamma=1/2$, $1$, and $2$, covering the whole range of expected $\Gamma$s in real gases and to compare with our numerical simulations, which also fall in this range.

\subsubsection{Density contrast for $\Gamma=1/2$ (soft EOS)}
Setting $\Gamma=1/2$ in Equation~(\ref{eq:nasty}) allows us to solve for the density contrast, which yields three formal solutions, but the only physical solution is
\begin{equation} \label{eq:solg1half}
x \equiv \frac{\rho}{\rho_0} = \frac{1}{8} \left(4\,b^2\mach^2 + b^4 \mach^4 + b^3 \mach^3 \sqrt{8 + b^2 \mach^2}\right).
\end{equation}
The trivial solution is $x=1$ (i.e., no density contrast) and the other non-trivial solution in this case fails to reproduce the boundary condition $x=1$ for $b\mach=1$, which must always be fulfilled when a shock just starts to form, i.e., $\mach_\parallel\to1$. Both these formal solutions are excluded, which leaves us with the only physical solution given by Equation~(\ref{eq:solg1half}) for the case $\Gamma=1/2$.

\subsubsection{Density contrast for $\Gamma=1$ (isothermal EOS)}
For $\Gamma=1$ in Equation~(\ref{eq:nasty}) we find two formal solutions, with the only non-trivial one being
\begin{equation} \label{eq:solg1}
x \equiv \frac{\rho}{\rho_0} = b^2\mach^2,
\end{equation}
which is the well-known solution for isothermal gas, as derived before by \citet{PadoanNordlund2011} and \citet{MolinaEtAl2012}. Thus, our generalised Equation~(\ref{eq:nasty}) for the density contrast naturally includes the trivial case of isothermal gas.

\subsubsection{Density contrast for $\Gamma=2$ (stiff EOS)}
The only physical solution of Equation~(\ref{eq:nasty}) for $\Gamma=2$ is
\begin{equation} \label{eq:solg2}
x \equiv \frac{\rho}{\rho_0} = \frac{1}{2} \left(-1 + \sqrt{1 + 8\,b^2\mach^2}\right).
\end{equation}

\subsection{Theoretical prediction for the density variance -- Mach number relation}

Now that we have derived the density contrast $\rho/\rho_0$ in non-isothermal, polytropic shocks for three extreme cases, $\Gamma=1/2$, $1$, and $2$ given by Equations~(\ref{eq:solg1half}), (\ref{eq:solg1}), and~(\ref{eq:solg2}), respectively, we can now insert these solutions into the density variance -- Mach number relation, Equation~(\ref{eq:molina_approx}), which immediately yields $\sigma_{\rho/\rho_0}^2$ as a function of the turbulence parameters $b$ and $\mach$ for each $\Gamma$. We do not repeat the corresponding solutions for $\sigma_{\rho/\rho_0}^2$ here, because they are simply given by the density contrast itself. Instead, we apply the standard conversion from linear density variance $\sigma_{\rho/\rho_0}^2$ to logarithmic density variance $\sigs^2$ in the variable $s=\ln(\rho/\rho_0)$, which---independent of the underlying distribution---is always given by
\begin{align} 
\sigs^2 & = \ln\left(1+\sigma_{\rho/\rho_0}^2\right)  \\
 & \simeq \ln\left(1+\rho/\rho_0\right) \label{eq:sigsconversion},
\end{align}
as routinely used, because the PDF of the logarithmic density contrast $s$ is nearly lognormal for the case $\Gamma=1$ \citep{PadoanNordlundJones1997,PassotVazquez1998,FederrathKlessenSchmidt2008,PriceFederrathBrunt2011,MolinaEtAl2012}. We follow the same definitions in order to enable direct comparisons with these previous works. Although the PDF for non-isothermal gas ($\Gamma\neq1$) is not lognormal, as we have seen in Figure~\ref{fig:pdfs}, we can still use the same definitions. Inserting our solutions for the density contrast from Equations~(\ref{eq:solg1half}), (\ref{eq:solg1}), and~(\ref{eq:solg2}) into Equation~(\ref{eq:sigsconversion}) yields the following new density variance -- Mach number relations:
\begin{align}
& \sigs^2 = \ln\bigg[1+\frac{1}{8} \left(4\,b^2\mach^2 + b^4 \mach^4 + b^3 \mach^3 \sqrt{8 + b^2 \mach^2}\right)\!\bigg] \nonumber \\
& \quad\text{for}\;\Gamma=1/2, \label{eq:solsigs1half} \\
& \sigs^2 = \ln\bigg[1+b^2\mach^2\bigg] \quad\text{for}\;\Gamma=1, \label{eq:solsigs1} \\
& \sigs^2 = \ln\bigg[1+\frac{1}{2} \left(-1 + \sqrt{1 + 8\,b^2\mach^2}\right)\!\bigg] \quad\text{for}\;\Gamma=2. \label{eq:solsigs2}
\end{align}

Finally, these relations can be modified to account for magnetic pressure, by replacing the thermal pressure $P$ in the derivation, Equation~(\ref{eq:rh2}), with the sum of the thermal and magnetic pressures:
\begin{align}
P & \to P + P_\mathrm{mag} \nonumber \\
\iff \Gamma\rho\cs^2 & \to \Gamma\rho\cs^2 + (1/2)\rho\va^2 \label{eq:magextension}.
\end{align}
Using the square of the Alfv\'en speed, $\va^2=2\cs^2\beta^{-1}$ from the standard definition of the plasma $\beta=P/P_\mathrm{mag}$, we can replace the sound speed by an effective magnetic sound speed and the Mach number by an effective magnetic Mach number,
\begin{align}
\cs & \to \cs\left(1+\beta^{-1}\right)^{1/2} \label{eq:csmag}, \\
\mach & \to \mach\left(1+\beta^{-1}\right)^{-1/2} \label{eq:machmag}.
\end{align}
Replacing $\mach$ in Equations~(\ref{eq:solsigs1half}), (\ref{eq:solsigs1}), and~(\ref{eq:solsigs2}) accounts for magnetic pressure, which stiffens the gas upon compression and reduces the density variance with respect to the non-magnetised case, because of the additional magnetic pressure. We note that the simple replacement of the sonic Mach number given by Equation~(\ref{eq:machmag}) is equivalent to the more elaborate derivations presented in \citet{PadoanNordlund2011} and \citet{MolinaEtAl2012} and yields the same replacement formula as previously derived for purely isothermal gas in \citet{FederrathKlessen2012}, because $\Gamma$ cancels out during the replacement steps above.

Thus, we have derived theoretical predictions for the density variance, as a function of $\mach$, $b$, $\beta$, and $\Gamma$, which we will now compare to the results of our numerical simulations, in order to test these predictions.

\subsection{Theory--simulation comparison of the $\sigs$--$\mach$ relation in polytropic gases}

\begin{figure*}
\centerline{\includegraphics[width=0.75\linewidth]{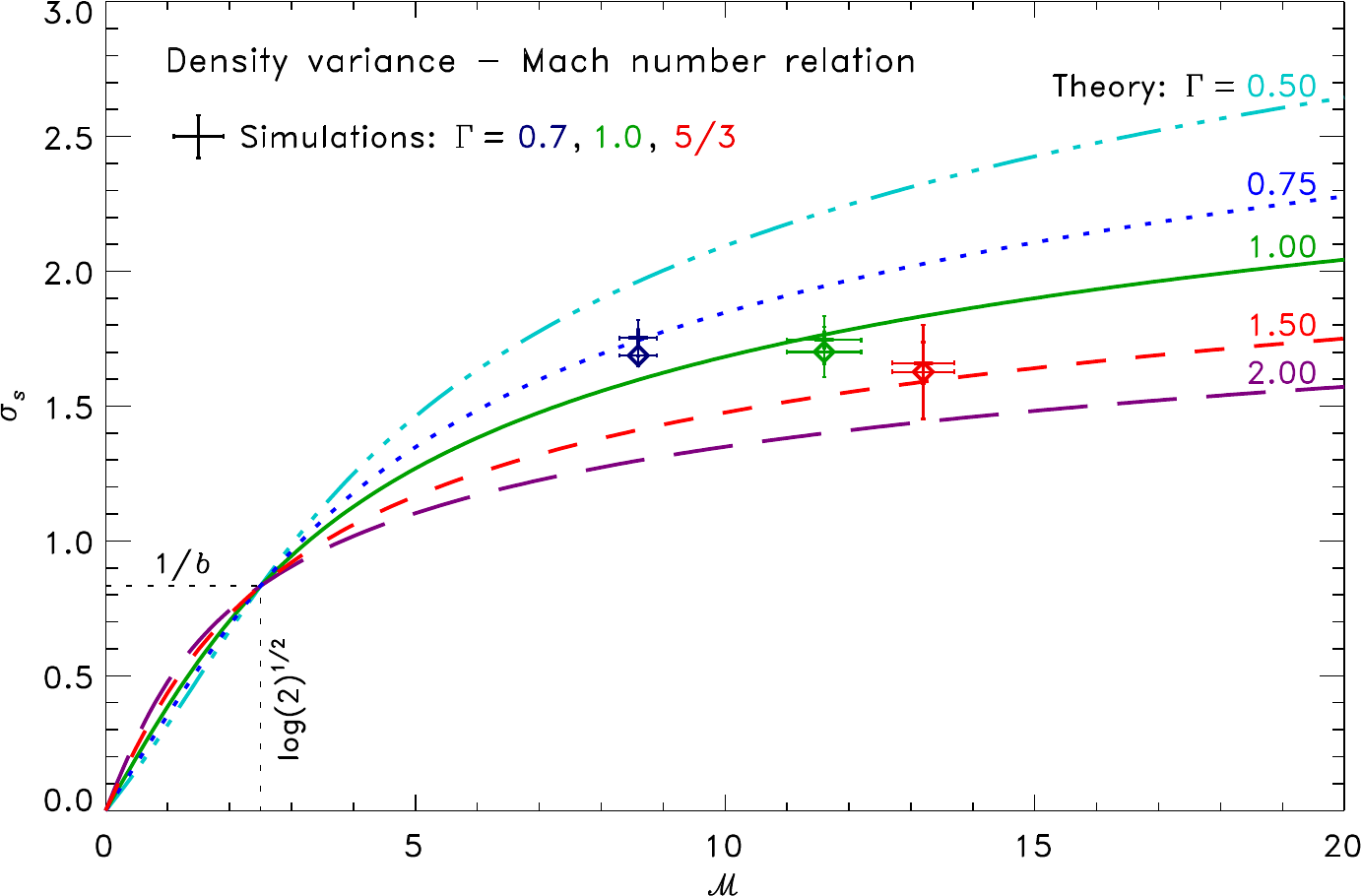}}
\caption{Density variance -- Mach number relation for the simulations and for the theoretical predictions based on Equation~(\ref{eq:nasty}). Simulation data points are shown for $\Gamma=0.7$ (blue), $1.0$ (green), and $5/3$ (red), where the crosses and diamonds respectively correspond to $\sigs=\sqrt{\sigsv\sigsm}$ directly taken from the simulation data and reconstructed based on the \citet{Hopkins2013PDF} fit to the density PDFs shown in Figure~\ref{fig:pdfs}. They agree within the 1-sigma uncertainties shown as error bars for each simulation datapoint (cf.~Table~\ref{tab:sims}). Theoretical curves are shown for $\Gamma=0.5$, $0.75$, $1$, $1.5$, and $2$. The analytic solutions for $\Gamma=0.5$, $1$, and $2$ correspond to Equations~(\ref{eq:solsigs1half}), (\ref{eq:solsigs1}), and~(\ref{eq:solsigs2}), respectively. The intermediate curves for $\Gamma=0.75$ and $1.5$ are numerical solutions of Equation~(\ref{eq:nasty}). The theory and simulations agree within the error bars. The intersection of each theoretical curve is always located at $\mach_\times=1/b$ and $\sigs{_{,\times}}=\sqrt{\log(2)}$, independent of $\Gamma$. Note that all theory curves and simulations use $b=0.4$, which corresponds to the natural mode mixture produced by the turbulent driving applied here \citep{FederrathDuvalKlessenSchmidtMacLow2010}.}
\label{fig:sigs}
\end{figure*}

Figure~\ref{fig:sigs} shows the theoretical $\sigs$--$\mach$ relations derived in the previous section, for $\gamma=0.5$, $0.75$, $1$, $1.5$, and $2$. The analytic solutions for $\Gamma=0.5$, $1$, and $2$ correspond to Equations~(\ref{eq:solsigs1half}), (\ref{eq:solsigs1}), and~(\ref{eq:solsigs2}), respectively, while the theoretical curves for $\Gamma=0.75$ and $1.5$ where obtained by numerical integration of Equation~(\ref{eq:nasty}).

First of all, we see that the density variance decreases with increasing polytropic $\Gamma$. This is expected, because increasing $\Gamma$ leads to higher pressure in the shocks, stopping them from becoming denser. We also saw in Figure~\ref{fig:imagesproj} that density fluctuations are smoothed when $\Gamma$ is increased. Both lead to a decreasing $\sigs$ with increasing $\Gamma$.

We now add our numerical simulations to the theoretical curves in Figure~\ref{fig:sigs}. They are shown as crosses and diamonds with the 1-sigma uncertainties plotted as error bars. The simulation data agree very well with the theoretical prediction. Minor deviations come from the fact that we would have to compute the rms \emph{pre-shock} Mach number in the 3D simulations for this theory--simulation comparison. This is because the pre-shock Mach number determines the density contrast in our theoretical derivation. Thus, we would have to detect the pre-shock gas and compute the rms Mach number only from that gas. In Figure~\ref{fig:sigs}, instead of the pre-shock Mach number, we plot the volume-weighted rms Mach number averaged over \emph{all} the gas, including contributions from the post-shock gas. However, the post-shock gas is primarily located in the dense shocks, by definition, so we can reasonably approximate the pre-shock Mach number by taking the volume-weighted rms Mach number shown in panel~c of Figure~\ref{fig:tevol}. The volume-weighted quantities primarily correspond to pre-shock gas, because most of the volume is pre-shock gas, while most of the mass is in post-shock gas. For the same reason, it is also important to get a global value of the density variance $\sigs$ from the simulations. The theories indeed predict $\sigs$ for \emph{all} the gas, including both pre-shock and post-shock contributions to the total variance. This is why we have to take both the volume- and mass-weighted density variance into account for the comparison with the theoretical model of the $\sigs$--$\mach$ relation. The most straightforward combination is an arithmetic or geometric mean of $\sigsv$ and $\sigsm$. Fortunately, it turns out that both arithmetic and geometric mean are very similar.

In summary, Figure~\ref{fig:sigs} shows very good agreement between the theoretical prediction of the density variance -- Mach number relation with the simulations. In this comparison, we have to be careful to evaluate the Mach number and the density variance for the appropriate shock regions. The theory provides an average of the pre-shock and post-shock density variance as a function of only the pre-shock Mach number. Thus, we approximate the \emph{total} density variance as the mean of the volume- and mass-weighted variance, $\sigs=\sqrt{\sigsv\sigsm}$ and we approximate the pre-shock Mach number with the volume-weighted rms Mach number in the simulations. This yields excellent agreement between our new theoretical $\sigs$--$\mach$ relations and the numerical simulations of polytropic turbulence.

\section{The star formation rate of polytropic turbulence} \label{sec:sfr}

Here we derive a theoretical prediction for the dependence of the SFR on the polytropic exponent $\Gamma$. We first briefly review previous results for isothermal gas ($\Gamma=1$) based on a simple lognormal approximation of the density PDF and then generalise the basic ansatz for the SFR to non-lognormal PDFs arising in highly intermittent and non-isothermal gas ($\Gamma\neq1$).

\subsection{The SFR in isothermal gas ($\Gamma=1$)}

Our starting point is the summary of SFR models in \citet{FederrathKlessen2012} and \citet{PadoanEtAl2014}, which are all based on the statistics of supersonic self-gravitating turbulence. The simple idea behind this derivation is that only dense gas above a certain density threshold (to be determined in Section~\ref{sec:scrit}) forms stars. Thus, we just have to integrate the density PDF from the threshold to infinity, weighted by $\rho/\tff(\rho)$ with the freefall time $\tff(\rho)=(3\pi/32G\rho)^{1/2}$, in order to derive an $\sfr$, i.e., the mass of a cloud forming stars per unit time,
\begin{equation} \label{eq:sfrbasic}
\sfr \sim \mathlarger{\int}_{\rhocrit}^{\infty}{\frac{\rho}{\tff(\rho)}\,p(\rho)\,\deriv \rho}.
\end{equation}
This integral can be written in terms of the logarithmic density $s\equiv\ln(\rho/\rho_0)$ to simplify the integration and to enable us to use the standard normalised form of the density PDF $p_V(s)$, as for example plotted in Figure~\ref{fig:pdfs},
\begin{equation} \label{eq:sfrmain}
\sfr \sim \mathlarger{\int}_{\scrit}^{\infty}{\exp\left(\frac{3}{2}s\right)\,p_V(s)\,\deriv s}.
\end{equation}
Note that the coefficient $3/2$ in the exponential term comes from the transformation of $\rho/\tff(\rho)\sim\rho/\rho^{-1/2}\sim\rho^{3/2}\sim\exp(3s/2)$. Equation~(\ref{eq:sfrmain}) is known as the \emph{multi-freefall model} of the SFR, because the density-dependence of the freefall time is evaluated \emph{inside} the integral \citep{HennebelleChabrier2011,FederrathKlessen2012}. Assuming a lognormal PDF given by Equation~(\ref{eq:lognormalpdf}), we can solve Equation~(\ref{eq:sfrmain}) analytically \citep{FederrathKlessen2012}, resulting in
\begin{equation} \label{eq:sfr_isothermal}
\sfr \sim \frac{1}{2} \exp\left(\frac{3}{8}\sigsv^2\right) \left[1+\mathrm{erf}\left(\frac{\sigsv^2-\scrit}{\left(2\sigsv^2\right)^{1/2}}\right)\right].
\end{equation}
For this result to be useful, we still have to specify the volume-weighted density variance $\sigsv^2$ and the threshold density $\scrit$, which we will do after we have derived the SFR for the general case where $\Gamma\neq1$, in the next section. We emphasise that we cannot just insert our new density variance -- Mach number relations for $\Gamma\neq1$ from Equations~(\ref{eq:solsigs1half})--(\ref{eq:solsigs2}) here in this SFR form, because Equation~(\ref{eq:sfr_isothermal}) was derived by assuming that the PDF is lognormal, which is a bad approximation for $\Gamma\neq1$ (cf.~Figure~\ref{fig:pdfs}). We have to use a better form of the PDF in cases with $\Gamma\neq1$.

\subsection{The SFR in non-isothermal gas ($\Gamma\neq1$)}

The solution for the SFR given by Equation~(\ref{eq:sfr_isothermal}) is strictly valid only for an exactly lognormal PDF. However, as we have seen in Figure~\ref{fig:pdfs}, a lognormal approximation for the density PDF in cases with $\Gamma\neq1$ is not appropriate and even for isothermal gas ($\Gamma=1$), the PDFs are subject to intermittency corrections, causing them to depart from the simple lognormal form, due to skewness and kurtosis \citep{Vazquez1994,PadoanNordlundJones1997,PassotVazquez1998,LiKlessenMacLow2003,KritsukEtAl2007,KowalLazarianBeresnyak2007,FederrathDuvalKlessenSchmidtMacLow2010,PriceFederrath2010,KonstandinEtAl2012ApJ,Hopkins2013PDF,Federrath2013}. Especially strong compressibility induced by hypersonic shocks and compressive driving of the turbulence leads to intense intermittent fluctuations \citep{FederrathDuvalKlessenSchmidtMacLow2010,Federrath2013}. Thus, we have to solve the general ansatz for the SFR given by Equation~(\ref{eq:sfrmain}) with a more appropriate form of the PDF $p_V(s)$. As we have seen in Figure~\ref{fig:pdfs}, the density PDFs for all cases ($\Gamma=1$ and $\Gamma\neq1$) can be well approximated with the intermittency PDF model by \citet{Hopkins2013PDF}, given by Equation~(\ref{eq:hopkinspdf}), so we insert that $p_V(s)$ into Equation~(\ref{eq:sfrmain}). The resulting integral cannot be solved analytically anymore, so we have to resort to semi-analytic solutions. To this end, we have to determine the two \citet{Hopkins2013PDF} PDF parameters, $\theta$ and $\sigsv$.

\subsubsection{Relation for the intermittency parameter $\theta$}

\begin{figure}
\centerline{\includegraphics[width=0.97\linewidth]{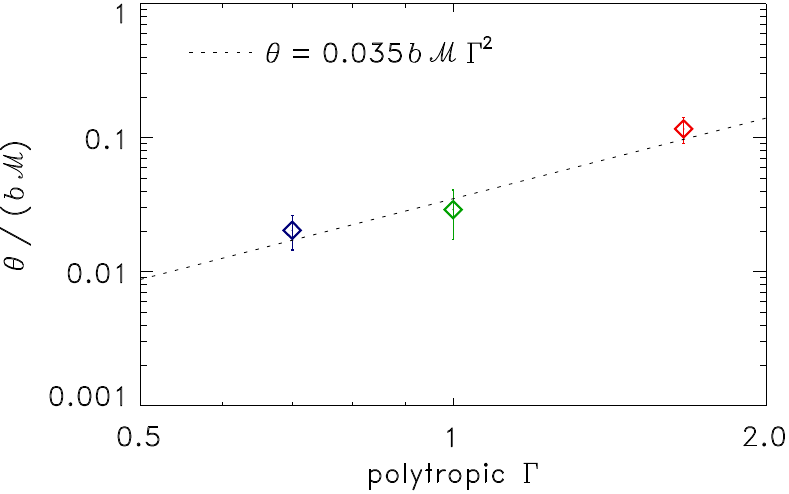}}
\caption{The \citet{Hopkins2013PDF} PDF intermittency parameter $\theta$ as a function of polytropic $\Gamma$ in our simulations. The dotted line shows a fit with $\theta = 0.035\,b\,\mach\,\Gamma^2$, which provides a reasonably good approximation for the dependence of $\theta$ on the fundamental turbulence parameters, $b$, $\mach$, and $\Gamma$.}
\label{fig:thetafit}
\end{figure}

First, we need a relation between the intermittency parameter $\theta$ of the \citet{Hopkins2013PDF} PDF and turbulence parameters, such as the Mach number $\mach$ and the driving mixture $b$. Fortunately, \citet{Hopkins2013PDF} already established such a relation and found that $\theta\sim b\mach$ for $\Gamma=1$ with a proportionality constant of $\sim0.05$. Here we extend this relation for cases where $\Gamma\neq1$, by fitting to our simulation dataset (cf.~the last column of Table~\ref{tab:sims}). Figure~\ref{fig:thetafit} shows the intermittency parameter measured in our simulations from the \citet{Hopkins2013PDF} density PDFs of Figure~\ref{fig:pdfs} as a function of $\Gamma$. We see a clear trend of increasing intermittency $\theta$ with increasing $\Gamma$, roughly following a power law given by
\begin{equation} \label{eq:theta}
\theta = 0.035\,b\,\mach\,\Gamma^2.
\end{equation}
Note that we find a somewhat smaller coefficient of $0.035$ than \citet{Hopkins2013PDF}, but it is still within the uncertainties between our and Hopkins' fit for the special case $\Gamma=1$. Equation~(\ref{eq:theta}) is the first required link between the \citet{Hopkins2013PDF} PDF and the fundamental cloud parameters $b$, $\mach$, and $\Gamma$, which determine the SFR.

\subsubsection{Relation for the density variance $\sigsv^2$}

Second, we have to find a relation for the volume-weighted density variance $\sigsv^2$. Fortunately, we have just derived new expressions for this as a function of $\Gamma$ in Section~\ref{sec:densityvariancemachrelation} and Figure~\ref{fig:sigs}. We must be careful, however, because we have seen in Section~\ref{sec:densityvariancemachrelation} that the derived Equations~(\ref{eq:solsigs1half})--(\ref{eq:solsigs2}) actually provide a \emph{combination of volume-weighted and mass-weighted density variance} and not directly the volume-weighted $\sigsv$, which is what we actually have to insert into the \citet{Hopkins2013PDF} PDF. Furthermore, we found in Section~\ref{sec:densityvariancemachrelation} and in Figure~\ref{fig:sigs} that the total derived $\sigs$ can be well approximated as the average (either geometric or arithmetic mean) of $\sigsv$ and $\sigsm$. Using the geometric mean for simplicity and inserting the relation between $\sigsv^2$ and $\sigsm^2$ from Equation~(\ref{eq:sigsvsigsm}), we find
\begin{align}
\sigs^2 & \simeq \sigsv\sigsm = \sigsv^2\left(1+\theta\right)^{-3/2} \nonumber \\
& \iff \sigsv^2 = \sigs^2 \left(1+\theta\right)^{3/2} \nonumber \\
& \iff \sigsv^2 = \sigs^2 \left(1+0.035\,b\,\mach\,\Gamma^2\right)^{3/2}, \label{eq:sigsintermittencycorrection}
\end{align}
where we used Equation~(\ref{eq:theta}) in the last step. We emphasise that this relation is only relevant for intermittency $\theta>0$. In the well-known and previously studied case of ideal isothermal and non-intermittent turbulence, the PDF is lognormal and the volume-weighted $\sigsv$ and the mass-weighted $\sigsm$ are identical \citep[e.g.,][]{Vazquez1994,LiKlessenMacLow2003}. That special case is included by our general Equation~(\ref{eq:sigsintermittencycorrection}) as the limiting case with zero-intermittency ($\theta=0$), for which indeed $\sigs=\sigsv=\sigsm$. However, we have to account for the fact that generally $\sigsv\neq\sigsm$, due to skewness and kurtosis in the PDF (cf.~\ref{fig:pdfs}). This is what we achieve with the new relation established in Equation~(\ref{eq:sigsintermittencycorrection}). With Equation~(\ref{eq:sigsintermittencycorrection}) in hand, we can now directly use our new density variance -- Mach numbers relations for $\sigs^2$ (Equations~\ref{eq:solsigs1half}--\ref{eq:solsigs2}) from Section~\ref{sec:densityvariancemachrelation} in order to get $\sigsv^2$ as a function of the basic cloud parameters $\Gamma$, $\mach$, and $b$.

\subsubsection{The density threshold for star formation} \label{sec:scrit}

Finally, we need a model for the density threshold $\scrit$, which serves as the lower limit of the SFR integral in Equation~(\ref{eq:sfrmain}). Models for $\scrit$ based on the \citet{KrumholzMcKee2005}, \citet{PadoanNordlund2011} and \citet{HennebelleChabrier2011} theories were already discussed in \citet{FederrathKlessen2012} and \citet{PadoanEtAl2014}. For the sake of simplicity and because we are here primarily interested in how the SFR depends on $\Gamma$, we ignore magnetic fields and use the critical density of the \citet{KrumholzMcKee2005} and \citet{PadoanNordlund2011} models, which are identical in this case. They were furthermore found to provide the best prediction of the SFR in star cluster formation simulations by \citet{FederrathKlessen2012} and \citet{FederrathEtAl2014}. Thus, the density threshold is a result of turbulence balancing gravity at the sonic scale \citep[e.g.,][]{VazquezBallesterosKlessen2003,McKeeOstriker2007,FederrathDuvalKlessenSchmidtMacLow2010}, which leads to
\begin{equation} \label{eq:scrit}
\scrit \sim \ln\left(\alphavir\mach^2\right),
\end{equation}
where $\alphavir=2E_\mathrm{kin}/E_\mathrm{grav}$ is the ratio of twice the kinetic to gravitational energy of a cloud, known as the virial parameter \citep{BertoldiMcKee1992,FederrathKlessen2012}. The coefficients in Equation~(\ref{eq:scrit}) are of order unity and were determined in \citet{FederrathKlessen2012}, but for our purposes, it is sufficient to consider only the basic dependence of $\scrit$ on $\alphavir$ and $\mach$.

\subsection{Theoretical prediction of the SFR as a function of $\Gamma$}

Now that we have $\theta$, $\sigsv^2$ and $\scrit$ as a function of $\Gamma$, $\mach$, $b$, and $\alphavir$ from Equations~(\ref{eq:theta})--(\ref{eq:scrit}), we can directly insert them together with the \citet{Hopkins2013PDF} PDF (Equation~\ref{eq:hopkinspdf}) into the main SFR Equation~(\ref{eq:sfrmain}), which leads to the following symbolic form:
\begin{align}
\begin{multlined} \label{eq:sfrsybolicsolution}
	\sfr(\Gamma,\alphavir,\mach,b) \sim \phantom{\mathlarger{\int}_{\scrit(\alphavir,\mach)}^{\infty}} \\
\mathlarger{\int}_{\scrit(\alphavir,\mach)}^{\infty}{e^{3s/2}\,p_V(s,\sigsv^2(\Gamma,\mach,b),\theta(\Gamma,\mach,b))\,\deriv s}.
\end{multlined}
\end{align}
The dependence on $\Gamma$ enters in the PDF $p_V$ through the volume-weighted density variance $\sigsv^2(\sigs^2(\Gamma,\mach,b),\Gamma,\mach,b)$ (Equation~\ref{eq:sigsintermittencycorrection}), through our new derivation of the density variance -- Mach number relation for $\sigs^2(\Gamma,\mach,b)$ from Section~\ref{sec:densityvariancemachrelation}, and through the dependence of the intermittency parameter $\theta(\Gamma,\mach,b)$ via Equation~(\ref{eq:theta}).

We can now go ahead and solve Equation~(\ref{eq:sfrsybolicsolution}) numerically. To keep it simple and to focus on the dependence of the SFR on $\Gamma$, we choose standard Milky Way cloud parameters and fix them to $\mach=10$ \citep{Larson1981,ElmegreenScalo2004,MacLowKlessen2004,McKeeOstriker2007,SchneiderEtAl2012,HennebelleFalgarone2012,Federrath2013sflaw}, $b=0.4$ \citep{FederrathKlessenSchmidt2008,FederrathDuvalKlessenSchmidtMacLow2010,BruntFederrathPrice2010a,BruntFederrathPrice2010b,Brunt2010,PriceFederrathBrunt2011,BurkhartLazarian2012,KainulainenTan2013,KainulainenFederrathHenning2013}, and $\alphavir=1$ \citep[][]{Larson1981,HeyerEtAl2009,KauffmannEtAl2013}. We then solve the coupled system of equations for a range of polytropic exponents, \mbox{$\Gamma=0.1$--$1.9$} in steps of $\Delta\Gamma=0.1$. A Mathematica\texttrademark{} notebook, which combines all the relevant equations and solves for the density contrast and for the SFR is available from the authors.

\begin{figure*}
\centerline{\includegraphics[width=0.75\linewidth]{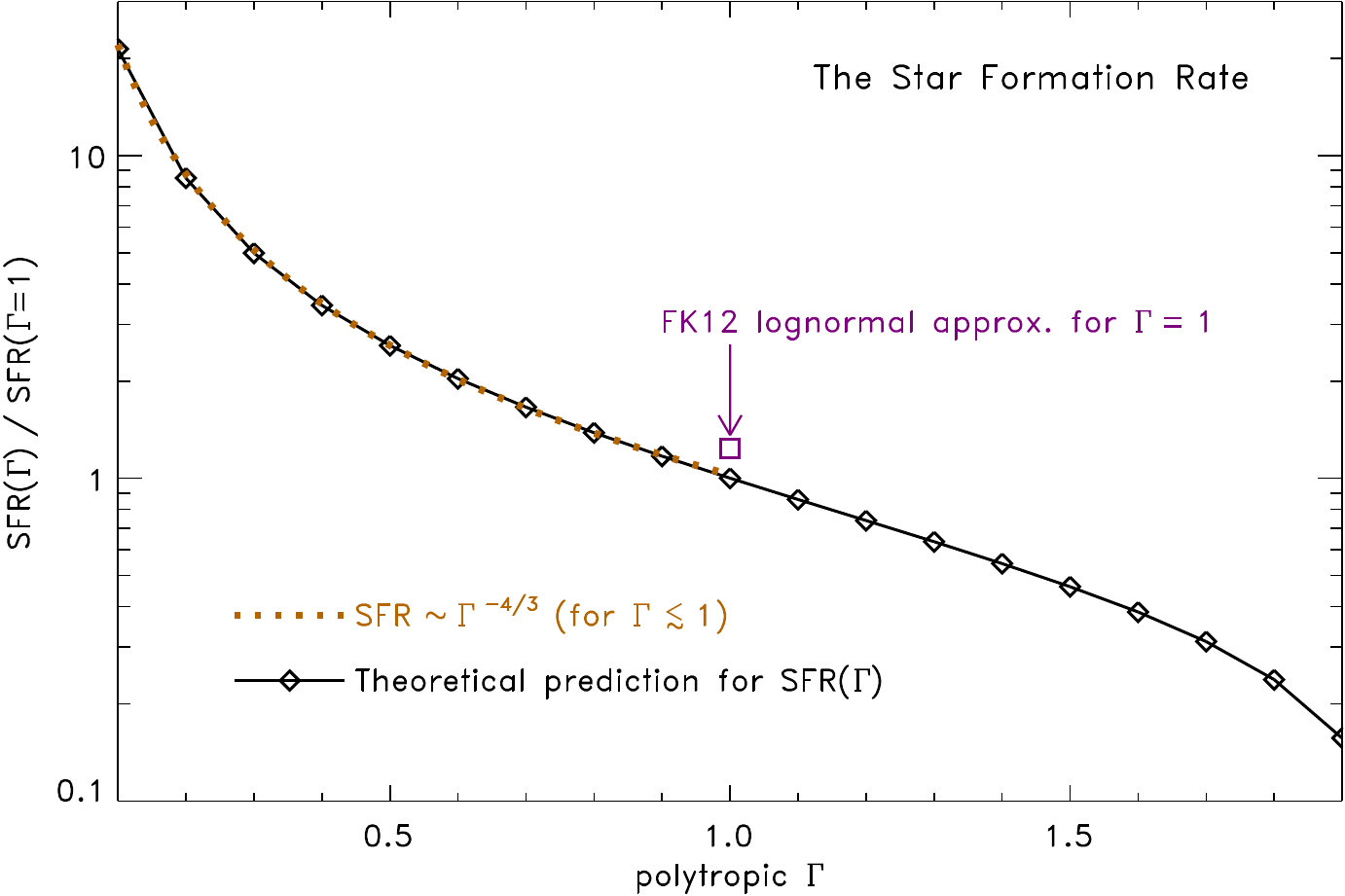}}
\caption{Dependence of the SFR on the polytropic exponent $\Gamma$. The diamonds are semi-analytic solutions of Equation~(\ref{eq:sfrsybolicsolution}) for \mbox{$\Gamma=0.1$--$1.9$} in steps of $\Delta\Gamma=0.1$ and standard parameters ($\mach=10$, $b=0.4$, $\alphavir=1$), using an integral over the \citet{Hopkins2013PDF} PDF. The dotted line is a power-law approximation with $\sfr\sim\Gamma^{-4/3}$ valid for $\Gamma\lesssim1$. The box shows the corresponding $\sfr$ computed with the lognormal PDF approximation for $\Gamma=1$ from FK12 \citep{FederrathKlessen2012}, which is strictly valid only for zero-intermittency ($\theta=0$). Fortunately, the intermittency correction introduced with the \citet{Hopkins2013PDF} PDF is relatively small for $\Gamma=1$, reducing $\sfr$ by $\sim19\%$ compared to the simple lognormal approximation. For $\Gamma\neq1$ however, the lognormal approximation breaks down (cf.~Figure~\ref{fig:pdfs}), requiring us to integrate the \citet{Hopkins2013PDF} PDF in order to compute semi-analytic solutions for $\Gamma\neq1$ (shown as diamonds).}
\label{fig:sfr}
\end{figure*}

The result is plotted in Figure~\ref{fig:sfr}, which shows the SFR as a function of $\Gamma$. First of all, we compare the semi-analytic solution (shown as diamonds) provided by Equation~(\ref{eq:sfrsybolicsolution}) for the special case $\Gamma=1$ (isothermal gas) with the analytic solution for the same case (shown as a square). This is the only case where an analytic solution given by Equation~(\ref{eq:sfr_isothermal}) can be derived, assuming that the PDF is a lognormal distribution. We find that the more accurate semi-analytic integral over the \citet{Hopkins2013PDF} PDF  instead of the lognormal PDF gives a 19\% lower SFR than the lognormal approximation for $\Gamma=1$. This is because the \citet{Hopkins2013PDF} PDF accounts for some small fraction of intermittency present even in the $\Gamma=1$ case (cf.~Figures~\ref{fig:pdfs} and~\ref{fig:thetafit}), which skews the high-density PDF tail to somewhat smaller densities and thus reduces the SFR by a small fraction compared to the lognormal approximation. However, the difference between the analytic (lognormal) and semi-analytic (Hopkins) integral for $\Gamma=1$ is only $19\%$. From this, we conclude that the analytic estimate based on the lognormal approximation as summarised in \citet{FederrathKlessen2012} is accurate to within a few tens percentile.

For the general case with $\Gamma\ne1$, however, we need the semi-analytic estimate shown as diamonds in Figure~\ref{fig:sfr}. We see that the SFR depends on $\Gamma$ and varies by about two orders of magnitude in the range \mbox{$\Gamma=0.1$--$1.9$}. The relevant range of $\Gamma$ for molecular clouds, however, is significantly narrower. The polytropic exponent can be approximated with $\Gamma=1$ (isothermal gas) over a wide range of number densities, from \mbox{$n\sim1$--$10^{10}\,\cm^{-3}$} with the temperature varying between $T\sim3\,\mathrm{K}$ and $T\sim10\,\mathrm{K}$ for solar metallicity gas \citep{OmukaiEtAl2005}. Radiation-hydrodynamical calculations including chemical evolution and cooling by \citet{MasunagaInutsuka2000} also show that $\Gamma\sim1$ for $n\lesssim10^9\,\cm^{-3}$ and then it rises to $\Gamma\sim1.1$ for $10^9\lesssim n/\cm^{-3}\lesssim10^{11}$, $\Gamma\sim1.4$ for $10^{11}\lesssim n/\cm^{-3}\lesssim10^{16}$, followed by a phase where $\Gamma\sim1.1$ in which molecular hydrogen is dissociated ($10^{16}\lesssim n/\cm^{-3}\lesssim10^{21}$). Finally, the gas becomes almost completely optically thick ($\Gamma=5/3$), when the star is born ($n\gtrsim10^{21}\,\cm^{-3}$). It must be emphasised, however, that all the phases with $n\gtrsim10^{10}\,\cm^{-3}$ only occur inside the dense, collapsing cores with transonic to subsonic velocity dispersions \citep{GoodmanEtAl1998,MotteAndreNeri1998,JijinaEtAl1999,AndreEtAl2000,CaselliEtAl2002,CsengeriEtAl2011}, which have already decoupled from the large-scale, supersonic turbulence in the cloud. The phases with $\Gamma>1$ only apply in relatively high-density gas, which may affect only a very small fraction of the high-density tail in the PDF. These high-density corrections are thus not expected to change the overall cloud SFR significantly.

Densities around the peak of the PDF and higher are expected to primarily contribute to the SFR integral and it is indeed around such densities ($n\sim10^2$--$10^5\,\cm^{-3}$) that \citet{OmukaiEtAl2005} and \citet{GloverMacLow2007a,GloverMacLow2007b,GloverFederrathMacLowKlessen2010} find that the polytropic exponent can vary between $\Gamma\sim0.5$ and $1.1$, followed by the optically thick regime with $\Gamma=5/3$ at high densities. Looking at our semi-analytic predictions in Figure~\ref{fig:sfr}, we see that the SFR varies by about a factor of $\sim3$ in the range \mbox{$\Gamma=0.5$--$1.1$} and by a factor of $\sim5$ in the range \mbox{$\Gamma=0.7$--$5/3$}. We conclude that the dependence of the SFR on $\Gamma$ is significant, changing the SFR by factors of a few for solar-metallicity gas. The dependence of the SFR on $\Gamma$ may be even more important for low-metallicity gas or in extreme environments such as starburst galaxies, where the heating and cooling balance can lead to $\Gamma$-values significantly different from unity \citep{AbelBryanNorman2002,GreifEtAl2008,WiseTurkAbel2008,SchleicherEtAl2010,RomeoBurkertAgertz2010,ClarkEtAl2011,HoffmannRomeo2012,SchoberEtAl2012,SafranekShraderEtAl2012,LatifEtAl2013}.

\section{Conclusions} \label{sec:conclusions}

We determined the density PDF in hydrodynamical simulations of supersonic, non-isothermal, polytropic turbulence. We run hydrodynamical simulations with grid resolutions of up to $2048^3$ cells and with polytropic exponents $\Gamma=0.7$, $1$, and $5/3$, approximating the thermodynamical properties of gas in the interstellar medium and in molecular clouds, for various density regimes \citep{MasunagaInutsuka2000,OmukaiEtAl2005,GloverMacLow2007a,GloverMacLow2007b,GloverFederrathMacLowKlessen2010}. We determine the filamentary structure of polytropic turbulence, measure the density PDF, and provide theoretical predictions for the density variance -- Mach number relation and for the SFR as a function of $\Gamma$. We now list our detailed conclusions:

\begin{enumerate}

\item Non-isothermal polytropic turbulence produces a complex network of shocks and filaments. A soft EOS ($\Gamma<1$) leads to the typical temperature structure seen in molecular clouds: cold dense gas surrounded by warm diffuse gas. The filaments are more fragmented on small scales if $\Gamma<1$, while turbulent density fluctuations are smoothed out if $\Gamma>1$ (stiff EOS) (cf.~Figure~\ref{fig:imagesproj}).

\item Dense gas cools upon compression for $\Gamma<1$, leading to a lower sound speed in the shocks, while gas with $\Gamma>1$ heats up during compression, leading to an increased sound speed in the shocks and filaments (cf.~Figure~\ref{fig:imagesslice}).

\item It is important to distinguish volume-weighted and mass-weighted quantities for non-isothermal turbulence ($\Gamma\ne1$). For a fixed velocity dispersion, the volume-weighted rms Mach number increases with increasing $\Gamma$, while the mass-weighted rms Mach number decreases compared to isothermal gas (cf.~Figure~\ref{fig:tevol}).

\item The density PDF depends significantly on the polytropic exponent. For $\Gamma>1$ the PDF develops a power-law tail towards low densities, while it is close to a lognormal distribution for $\Gamma\lesssim1$ (cf.~Figure~\ref{fig:pdfs}). The variance and intermittency parameter of the density PDF are converged with numerical resolution.

\item The column-density PDFs (cf.~Figure~\ref{fig:pdfs_coldens}) show the same systematic trend with $\Gamma$ as the volumetric density PDFs. The standard deviation of the column-density contrast $\sigma_\eta$ produced in the simulations is consistent with observations.

\item Higher $\Gamma$ produces a wider pressure distribution than lower $\Gamma$. We find power-law tails towards low Mach number values in the Mach number PDFs, independent of $\Gamma$. The local Mach numbers reach several hundreds to a few thousand for $\Gamma=5/3$, while they are capped at a few tens for $\Gamma\lesssim1$ (cf.~Figure~\ref{fig:pdfsmisc}).

\item The Mach number -- density correlations in the simulations match the theoretical expectation given by $\mach\sim\rho^{(1-\Gamma)/2}$ (cf.~Figure~\ref{fig:2dpdfs}).

\item Our new theoretical derivation of the density variance -- Mach number relation in polytropic gases is well reproduced by the outcome of the numerical simulations (cf.~Figure~\ref{fig:sigs}). We find that the density variance decreases with increasing $\Gamma$ for a fixed pre-shock (or volume-weighted) Mach number.

\item The intermittency of the density PDF (which is a measure for how strongly the PDF departs from a simple lognormal distribution) increases with increasing $\Gamma$ (cf.~Figure~\ref{fig:thetafit}). We provide a fit function that describes the dependence of the intermittency parameter $\theta$ on the Mach number $\mach$, the turbulent driving parameter $b$, and the polytropic $\Gamma$, given by $\theta=0.035\,b\,\mach\,\Gamma^2$.

\item We derive a theoretical prediction for the dependence of the SFR on $\Gamma$, by numerically integrating the \citet{Hopkins2013PDF} intermittency PDF, Equation~(\ref{eq:hopkinspdf}). For isothermal gas ($\Gamma=1$), we find that the intermittency corrections reduce the SFR by $\sim19\%$ compared to the previously established lognormal approximation. For \mbox{$\Gamma\neq1$}, however, intermittency corrections are important and lead to significant changes in the SFR. We find that the SFR increases by a factor of $\sim1.7$ for $\Gamma=0.7$ compared to $\Gamma=1$. For $\Gamma=5/3$, the SFR decreases by a factor of $\sim3$ compared to $\Gamma=1$ (cf.~Figure~\ref{fig:sfr}). This leads to overall variations in the SFR by a factor of $\sim5$ within the range $0.7\leq\Gamma\leq5/3$.

\end{enumerate}

We conclude that temperature fluctuations can introduce significant variations in the density PDF and in the SFR. While molecular clouds can be approximated as being close to isothermal ($\Gamma=1$) over a wide range of densities, there are regimes in which the EOS turns from isothermal to soft with $\Gamma=0.7$, and then to a stiff EOS with $\Gamma=1.1$, followed by $\Gamma>1.4$ when the gas becomes optically thick in the dense star-forming cores \citep{OmukaiEtAl2005}. Our study demonstrates that we expect a systematic evolution of the density PDF and SFR as the gas evolves and passes through these different thermodynamic phases.

\section*{Acknowledgements}
We thank P.~Hennebelle and E.~V\'azquez-Semadeni for stimulating discussions on polytropic turbulence, and we thank the anonymous referee for a constructive and useful report. C.F.~acknowledges funding provided by the Australian Research Council's Discovery Projects (grants~DP130102078 and~DP150104329).
We gratefully acknowledge the J\"ulich Supercomputing Centre (grant hhd20), the Leibniz Rechenzentrum and the Gauss Centre for Supercomputing (grant pr32lo), the Partnership for Advanced Computing in Europe (PRACE grant pr89mu), and the Australian National Computing Infrastructure (grant ek9).
The software used in this work was in part developed by the DOE-supported Flash Center for Computational Science at the University of Chicago.

\end{document}